\documentclass[reprint,a4paper,australian,amsmath,aps,prd,twocolumn,eqsecnum,superscriptaddress,preprintnumbers]{revtex4-1}

\usepackage{braket}

\usepackage{cancel} 
\usepackage{times}
\usepackage{graphicx}
\usepackage{amssymb}
\usepackage{amsmath}
\usepackage{mathrsfs}
\usepackage{xcolor}
\usepackage{dcolumn}
\usepackage{bm}
\usepackage{yhmath}
\usepackage{hyperref}
\usepackage{booktabs}
\usepackage{subfigure}
\usepackage{color}
\usepackage{multirow}
\usepackage{dcolumn}
\usepackage[normalem]{ulem}
\usepackage[capitalise]{cleveref}
\crefrangelabelformat{equation}{(#3#1#4--#5#2#6)}

\newcommand{\rcite}[1]{\cite{#1}}
\newcommand{\rref}[1]{\rcite{#1}}
\newcommand{\refref}[1]{Ref.~\rcite{#1}}

\newcommand{\eref}[1]{Eq.~(\ref{#1})}
\newcommand{\sref}[1]{Sec.~(\ref{#1})}
\newcommand{\tref}[1]{Table~\ref{#1}}
\newcommand{\fref}[1]{Fig.~(\ref{#1})}

\newcommand{\nn}{\nonumber}

\newcommand{\dkt}{}
\newcommand{\dkf}{}
\newcommand{\dxt}{}
\newcommand{\dxf}{}

\newcommand{\UCAS}{School of Physical Sciences, University of Chinese Academy of \\ Sciences (UCAS), Beijing 100049, China}
\newcommand{\CSSM}{Special Research Centre for the Subatomic Structure
  of Matter (CSSM),\\Department of Physics, University of
  Adelaide, Adelaide, South Australia 5005, Australia}
\newcommand{\CoEPP}{ARC Centre of Excellence for Particle Physics at
  the Terascale (CoEPP),\\Department of Physics, University
  of Adelaide, Adelaide, South Australia 5005, Australia}

\begin{document}
\title{
  {Partial-wave mixing in Hamiltonian effective field theory}
}
%
%
%
%
\author{Yan Li}
\affiliation{\UCAS}
\author{Jia-Jun Wu}
\affiliation{\UCAS}
\author{Curtis D. Abell}
\affiliation{\CSSM}
\author{Derek B. Leinweber}
\affiliation{\CSSM}
\author{Anthony W. Thomas}
\affiliation{\CSSM}
\affiliation{\CoEPP}
%

\begin{abstract} 
  The spectrum of excited states observed in the finite volume of lattice QCD is governed by the
  discrete symmetries of the cubic group.  This finite group permits the mixing of orbital angular
  momentum quanta in the finite volume.  As experimental results refer to specific angular momentum
  in a partial-wave decomposition, a formalism mapping the partial-wave scattering potentials to
  the finite volume is required.  This formalism is developed herein for Hamiltonian effective
  field theory, an extension of chiral effective field theory incorporating the L\"uscher relation
  linking the energy levels observed in finite volume to the scattering phase shift.  The formalism
  provides an optimal set of rest-frame basis states maximally reducing the dimension of the
  Hamiltonian, and it should work in any Hamiltonian formalism.
  As a first example of the formalism's implementation, lattice QCD results for the
  spectrum of an isospin-2 $\pi\pi$ scattering system are analyzed to determine the $s$, $d$, and
  $g$ partial-wave scattering information.
\end{abstract}

\maketitle

%
\section{Introduction}

The established nonperturbative approach to understanding the
emergent phenomena of the relativistic quantum field theory of the
strong interactions, quantum chromodynamics (QCD), is the numerical
approach of lattice QCD.
While experiment probes QCD through infinite-volume scattering
observables such as the phase shift and inelasticity, 
the finite-volume and Euclidean-time aspects of the lattice
formulation render the accessible quantity to be the spectrum of
states in the finite-volume lattice.
For the case of elastic two-body scattering in the rest frame,
L\"{u}scher \rref{Luscher:1985dn,Luscher:1986pf,Luscher:1990ux} proved
that these observables are related by what is now known as
L\"{u}scher's formula.

Up to exponentially suppressed corrections, the proof shows that a
quantum field theoretic system can be reduced to a quantum mechanical
system with an effective potential.  The infinite-volume phase shift
and the finite-volume spectrum are then related by the potential
independent L\"{u}scher's formula.
An equivalent approach is Hamiltonian effective field theory (HEFT).
In the standard approach, a potential is parametrized, fit to the
finite-volume lattice QCD spectrum, and the  infinite-volume
phase shift is then derived from the fit.
HEFT was formulated in the baryon sector with reference to the Delta baryon resonance. 
In the simplest case, the scattering of a nucleon and a pion through an intermediate Delta baryon basis state was considered. 
Upon rearranging the formal equation arising from $\det(H-\lambda I)=0$, one can make contact with the established result of chiral perturbation theory, either in a finite volume or in the continuum. 
Indeed, the structure of the terms of the Hamiltonian is dictated by chiral perturbation theory. 
In the weakly interacting perturbative limit, HEFT reproduces chiral perturbation theory in a finite volume.  

The equivalence, up to exponentially suppressed corrections, between
L\"{u}scher's method and HEFT has been examined in detail for the
single partial-wave case in Ref.~\cite{Wu:2014vma}.
In other words, HEFT provides an alternate bridge connecting the
finite-volume spectrum and the infinite-volume phase shifts, but does
not change the fixed relationship between them.

HEFT also provides insight into the structure of the finite-volume
eigenstates. 
In solving the Hamiltonian eigenvalue equation for the finite-volume energy eigenstates, one obtains an eigenvector describing the composition of the eigenstates in terms of the basis states of the Hamiltonian matrix.
The energy eigenvector describes the contribution of each basis state in the Hamiltonian to the eigenstate.
When the regulator takes phenomenologically motivated forms associated with the finite size of the hadrons participating in the scattering, one gains insight into the manner in which the Hamiltonian model states are constructed by the basis states of the Hamiltonian~\cite{Wu:2016ixr}.  
For example, in the low-lying nucleon spectrum, scattering states excited by lattice QCD interpolating fields are described within the Hamiltonian effective field theory in terms of the same hadronic scattering degrees of freedom~\cite{Liu:2015ktc,Liu:2016uzk,Liu:2016wxq,Wu:2017qve}.

While the model-independent finite-volume quantization condition approach, based upon the L\"uscher method, has been extended to various cases, including asymmetric boxes \cite{Feng:2004ua,Li:2003jn,Li:2007ey}, moving frames \cite{Davoudi:2011md,Fu:2011xz,Gockeler:2012yj,Kim:2005gf,Leskovec:2012gb,Rummukainen:1995vs}, the multichannel case \cite{Bernard:2010fp,Guo:2012hv,He:2005ey,Hu:2016shf,Lage:2009zv,Li:2012bi}, nonzero spins \cite{Beane:2003da,Beane:2003yx,Briceno:2013bda,Meng:2003gm}, twisted-boundary conditions \cite{Bedaque:2004ax,Bedaque:2004kc,deDivitiis:2004kq,Sachrajda:2004mi}, and the multibody case \cite{Blanton:2019vdk,Briceno:2012rv,Briceno:2017tce,Bulava:2019kbi,Doring:2018xxx,Guo:2018ibd,Hammer:2017kms,Hammer:2017uqm,Hansen:2014eka,Hansen:2015zga,Hansen:2017mnd,Hansen:2019nir,Jackura:2019bmu,Mai:2019fba,Meng:2017jgx,Pang:2019dfe,Polejaeva:2012ut,Romero-Lopez:2019qrt}, HEFT is still in the early stages of development.

HEFT was first introduced in Ref.~\cite{Hall:2013qba} to study a $\Delta
\to N\pi$ system, and was developed further in a series of
works \cite{Hall:2014uca,Wu:2014vma,Liu:2015ktc,Liu:2016uzk,Liu:2016wxq,Wu:2016ixr,Wu:2017qve}.
It is worth mentioning that it is straightforward to include more channels in the HEFT framework \cite{Wu:2014vma,Liu:2015ktc,Liu:2016uzk,Liu:2016wxq,Wu:2016ixr,Wu:2017qve}; however, it has not been extended to the high partial-wave case yet.
Until now, HEFT has only been applied to single partial-wave cases, in which the high dimension of the Hamiltonian matrix can be significantly reduced through the consideration of $C_3(N)$ symmetry, {\it i.e.} the symmetry associated with summing the squares of three integers to the value $N$.

Once the mixing of higher partial waves is taken into account one must 
abandon the use of $C_3(N)$ and work with higher dimension matrices.  
The focus of this investigation is to create an optimal set of
rest-frame cubic-group basis states, maximally reducing the dimension
of the Hamiltonian and enabling the determination of several
partial-wave scattering parameters simultaneously.
The formalism developed herein for partial-wave mixing is easily generalized to the nonzero spin case. 
For the moving-frame case, a method based on the formalism here is under development and will be presented in a future work.
We also note that the formalism works not only in HEFT, but also in any Hamiltonian formalism, {\it e.g.}, the harmonic oscillator basis effective theory \cite{McElvain:2019ltw,Drischler:2019xuo}.

In Sec. \ref{sec:IFVSBN}, the infinite-volume and finite-volume
Hamiltonians are introduced and relations between infinite- and
finite-volume potentials are established.
Sec. \ref{sec:MPW} presents the new formalism for creating the
rest-frame cubic-group basis states required to accommodate
partial-wave mixing.
In Sec. \ref{sec:EIS}, lattice QCD results for isospin-2 $\pi\pi$
scattering \cite{Dudek:2012gj} are examined to illustrate the formalism in practice and
examine the consistency between the formalism created here and
L\"{u}scher's method.
Finally, results are summarized in Sec. \ref{sec:Sum}.

%
\section{Infinite- and Finite-Volume Hamiltonians}
\label{sec:IFVSBN}

\renewcommand{\dkt}[1]{\frac{d^3\mathbf{#1}}{(2\pi)^3}}
\renewcommand{\dkf}[1]{\frac{d^4 #1}{(2\pi)^4}}
\renewcommand{\dxt}[1]{d^3\mathbf{#1}~}
\renewcommand{\dxf}[1]{d^4#1~}

The Hamiltonian operator of a three-dimensional infinite-volume (IFV) system can be written in bra-ket notation as
\begin{align}
  \hat{H} = \hat{H}_0 + \hat{V} &= \int \dkt{k}\,h(k) \ket{\mathbf{k}}\bra{\mathbf{k}} \nonumber\\
                                &+ \int \dkt{p}\dkt{k} \,V(\mathbf{p},\mathbf{k}) \ket{\mathbf{p}}\bra{\mathbf{k}} \,,
\label{Eq:IFV}
\end{align}
with the conventions
\begin{align}
  \mathbf{1} = \int \dxt{x} \ket{\mathbf{x}}\bra{\mathbf{x}} =& \int \dkt{k} \ket{\mathbf{k}}\bra{\mathbf{k}}\,, \; \braket{\mathbf{x}|\mathbf{k}} = e^{i\mathbf{k}\cdot\mathbf{x}} \,,\nn\\
  \braket{\mathbf{y}|\mathbf{x}} = \delta^3(\mathbf{y-x})\,,& \;
                                   \braket{\mathbf{p}|\mathbf{k}} = (2\pi)^3\,\delta^3(\mathbf{p-k})\,,
\end{align}
where $h(k)$ and $V(\mathbf{p},\mathbf{k})$ are the total kinematic
energy and the momentum-dependent potential
of two particles, respectively.  Then from
\begin{equation}
  \bra{\mathbf{x}}\hat{V}\ket{\mathbf{k}} = \int \dxt{y}~V(\mathbf{x},\mathbf{y})	\braket{\mathbf{y}|\mathbf{k}} \,,
\end{equation}
we have
\begin{align}\label{eq:HEFT-IFVP}
  V(\mathbf{p},\mathbf{k}) &= \bra{\mathbf{p}}\hat{V}\ket{\mathbf{k}} \nn\\
                           &= \int   \dxt{x}\dxt{y}e^{-i(\mathbf{p}\cdot\mathbf{x}-\mathbf{k}\cdot\mathbf{y})}
                                     \,V(\mathbf{x},\mathbf{y}) \,, 
\end{align}
where $V(\mathbf{x},\mathbf{y})$ (instead of $V(x)\,\delta^3(\mathbf{x-y})$) is introduced to allow
for the consideration of nonlocal potentials.

Correspondingly, in the finite periodic volume (FV) where momenta are discrete, the Hamiltonian can
be written with bra-ket notation as follows:
\begin{align}\label{eq:HEFT-HFVBK}
  \hat{H}_L = \hat{H}_{0L} + \hat{V}_L 
&= \sum_{\mathbf{n}\in\mathbb{Z}^3} h\left(\frac{2\pi n}{L}\right) \ket{\mathbf{n}}\bra{\mathbf{n}} \nonumber\\
&+ \sum_{\mathbf{n}',\mathbf{n}\in\mathbb{Z}^3} V_L\left(\frac{2\pi\,\mathbf{n}'}{L},\,\frac{2\pi\,\mathbf{n}}{L}\right) \ket{\mathbf{n}'}\bra{\mathbf{n}} \,,
\end{align}
with the conventions
\begin{align}
  &\mathbf{1} = \int_{L^3} d^3\mathbf{x}\, \ket{\mathbf{x}}\bra{\mathbf{x}}\, , 
   \mathbf{1} = \sum_{\mathbf{n}\in\mathbb{Z}^3} \ket{\mathbf{n}}\bra{\mathbf{n}}\,,\; \braket{\mathbf{n}'|\mathbf{n}} 
              = \delta_{\mathbf{n}',\mathbf{n}}\,, \nn\\
  &\braket{\mathbf{y}|\mathbf{x}} = \sum_{\mathbf{n}\in\mathbb{Z}^3} \delta^3(\mathbf{y-x+n}L)\,, \;
                                   \braket{\mathbf{x}|\mathbf{n}} = L^{-3/2}\,e^{i\frac{2\pi}{L}\,\mathbf{n}\cdot\mathbf{x}} \,,
\end{align}
where the subscript $L$ denotes the periodic boundary condition of length $L$, and $\int_{L^3}
d^3\mathbf{x}$ denotes an integral over the first period of coordinate space.

The relationship between $V(\mathbf{p},\mathbf{k})$ of
Eq.~(\ref{Eq:IFV}) and
$V_L\left({2\pi\,\mathbf{n}'}/{L},\,{2\pi\,\mathbf{n}}/{L}\right)$
of Eq.~(\ref{eq:HEFT-HFVBK}) is obtained through the consideration of
the conventions,
\begin{align}
  \mathbf{1} &= \int \dkt{k} \ket{\mathbf{k}} \bra{\mathbf{k}} \nn \\
             &\to
\left ( \frac{2\pi}{L}\right )^3\, \sum_{\mathbf{n}\in\mathbb{Z}^3} 
\frac{1}{\left ( 2\pi \right )^3}
L^{3/2}\, \ket{\mathbf{n}}\bra{\mathbf{n}}\, L^{3/2} \, ,
\end{align}
where $\ket{\mathbf{k}} \to L^{3/2}\, \ket{\mathbf{n}}$ to ensure the
convention $\sum_{\mathbf{n}\in\mathbb{Z}^3}
\ket{\mathbf{n}}\bra{\mathbf{n}} = \mathbf{1}$.  Now,
\begin{align}
&\int \dkt{p}\dkt{k} \,V(\mathbf{p},\mathbf{k}) \ket{\mathbf{p}}\bra{\mathbf{k}} \nn \\
&\to \sum_{\mathbf{n}',\mathbf{n}\in\mathbb{Z}^3} \, 
\left ( \frac{2\pi}{L}\right )^6\,
\frac{1}{\left ( 2\pi \right )^6}\,
L^{3/2}\, \ket{\mathbf{n}'}\bra{\mathbf{n}} \, L^{3/2} \, \nn \\
&\quad\times V\left(\frac{2\pi\,\mathbf{n}'}{L},\,\frac{2\pi\,\mathbf{n}}{L}\right) \, , \\
&\equiv \sum_{\mathbf{n}',\mathbf{n}\in\mathbb{Z}^3} \, 
\ket{\mathbf{n}'}\bra{\mathbf{n}} \, 
V_L\left(\frac{2\pi\,\mathbf{n}'}{L},\,\frac{2\pi\,\mathbf{n}}{L}\right) \, ,
\end{align}
where $V_L = V / L^3$.

L\"{u}scher's formula is a one-to-one relation only in the simplest
cases.  In more general cases, most data fail to find the partners
required to apply L\"{u}scher's formula directly.  As a consequence, a
fitting process is necessary. In the normal L\"{u}scher approach,
e.g. \refref{Dudek:2012gj}, the phase shift is parametrized and
constrained by lattice results.  In contrast, the potential is
parametrized and constrained in this approach.

The phase shifts can be solved from the Lippmann-Schwinger equation
\begin{equation}\label{eq:LSE}
  \hat{T} = \hat{V} + \hat{V} (E-\hat{H}_0+i\varepsilon)^{-1}\, \hat{T} \,,
\end{equation}
where $E$ is the total energy of system in the rest frame.
With the partial-wave expansions
\begin{align}\label{eq:PWE}
  \bra{\mathbf{p}}\hat{V}\ket{\mathbf{k}} &=\sum_{l,m}\, v_l(p,k)\, Y_{lm}(\hat{\mathbf{p}})\,Y_{lm}^*(\hat{\mathbf{k}}) \,, \nn\\
  \bra{\mathbf{p}}\hat{T}\ket{\mathbf{k}} &=\sum_{l,m}\, t_l(p,k;E)\, Y_{lm}(\hat{\mathbf{p}})\,Y_{lm}^*(\hat{\mathbf{k}}) \,,
\end{align}
we have
\begin{equation}
  t_l(p,k;E) = v_l(p,k) + \int \frac{q^2\,dq}{(2\pi)^3}\, \frac{v_l(p,q)\,t_l(q,k;E)}{E-h(q)+i \varepsilon} \,.
\end{equation}
Then the phase shift is given by
\begin{equation}
  e^{2i\,\delta_l(E)} = 1- i\frac{\bar{q}^2}{4\pi^2}\left( \frac{dh(q)}{dq} \right)^{-1}\Big{|}_{q=\bar{q}}\,
  t_l(\bar{q},\bar{q};E) \,,
\end{equation}
where $\bar{q}$ is the on-shell momentum, i.e., $h(\bar{q})=E$.

With a suitable momentum cutoff, the dimension of the FV Hamiltonian of \eref{eq:HEFT-HFVBK}
remains too high to solve for the spectrum. Moreover, one encounters an overwhelming number of
degeneracies in the spectrum.  In the next section, a formalism providing an optimal set of
rest-frame basis states is developed.  The formalism maximally reduces the dimension of the
Hamiltonian, and thus resolves the degeneracy problem.

\section{Mixing between Partial Waves}
\label{sec:MPW}

From rotational invariance, the infinite-volume potential can be expanded as in \eref{eq:PWE},
\begin{equation}\label{eq:SFV-PW}
  V(\mathbf{p},\mathbf{k}) = \sum_{l,m}\, v_l(p,k)\, Y_{lm}(\hat{\mathbf{p}})\, Y_{lm}^*(\hat{\mathbf{k}}) \,.
\end{equation}
Then we will have
\begin{align}\label{eq:SFV-VLlm}
  \hat{V}_L &= \sum_{\mathbf{n}',\mathbf{n}} \sum_{l,m} \,L^{-3}\, v_l(k_{N'},k_{N})\, Y_{lm}(\hat{\mathbf{n}}')\, Y_{lm}^*(\hat{\mathbf{n}}) \ket{\mathbf{n}'}\bra{\mathbf{n}}\,, \nn\\
            &= \sum_{N',N}\sum_{l,m} \,\frac{1}{4\pi L^3}\,v_l(k_{N'},k_{N})\, \nonumber\\
            &\quad\times \left(\sum_{\hat{\mathbf{n}}'}\sqrt{4\pi}\, Y_{lm}(\hat{\mathbf{n}}')\ket{\mathbf{n}'}\right) \left(\sum_{\hat{\mathbf{n}}}\sqrt{4\pi}\, Y_{lm}^*(\hat{\mathbf{n}})\bra{\mathbf{n}}\right)\,, \nn\\
            &= \sum_{N',N}\sum_l \,\tilde{v}_l(k_{N'},k_{N})\, \sum_{m} \ket{N';l,m}\bra{N;l,m} \,,
\end{align}
where we use $N$ to represent $\mathbf{n}^2$, and we introduce
\begin{align}
  &\tilde{v}_l(k_{N'},k_{N}) = \frac{1}{4\pi L^3}\, v_l(k_{N'},k_{N})\,, \nonumber\\
  &\qquad k_{N} = \frac{2\pi\, \sqrt{N}}{L}  ~~~~\text{and}~~~~ \sum_{\hat{\mathbf{n}}} = \sum_{\mathbf{n}^2=N}\,,
\end{align}
to simplify the notation.  We also define
\begin{equation}\label{eq:SFV-defnlm}
  \ket{N;l,m} =
  \begin{cases}
    \sum_{\hat{\mathbf{n}}}\sqrt{4\pi}\, Y_{lm}(\hat{\mathbf{n}})\ket{\mathbf{n}} & N\neq0 \\
    \delta_{l,0} \ket{\mathbf{n}=(0,0,0)} & N=0
  \end{cases}
  \,,
\end{equation}

In fact, if we hold $k_{N}=2\pi \sqrt{N}/L$ fixed and let $L$ go to infinity, then $N$ will also go to
infinity.  Then the summation in \eref{eq:SFV-defnlm} will be approximately proportional
to an integral over solid angle, and hence, $\ket{N;l,m}$ will be approximately proportional to an IFV
form, ($N_k\equiv\left(\frac{kL}{2\pi}\right)^2$)
\begin{equation}\label{eq:IFVSP}
  \ket{N_k;l,m} := \int d\Omega_{\hat{\mathbf{k}}}\, Y_{lm}(\hat{\mathbf{k}})\ket{\mathbf{k}} \,.
\end{equation}

In the case of the infinite volume, with rotational invariance, we can label the energy spectrum in
a spin-$l$ representation of the orthogonal group $O(3)$, and the spectrum can be extracted from a
reduced Hamiltonian
\begin{align}
  \hat{H}_l &= \int \frac{k^2\,dk}{(2\pi)^3}\, h(k)\, \sum_{l,m} \ket{{N_k};l,m}\bra{N_k;l,m} \nonumber\\
  &\quad+ \int \frac{p^2\,dp}{(2\pi)^3}\,\frac{k^2\,dk}{(2\pi)^3}\,v_l(p,k)\, \sum_{l,m} \ket{N_p;l,m}\bra{N_k;l,m} \,.
\end{align}
When we make the system finite, however, we can only describe the energy spectrum in an irreducible
representation $\Gamma$ of the cubic group $O_h$.
But the $\Gamma$ is also constrained by the initial rotational-invariant potential and the structure
of the lattice momentum sphere.

As states with different $N$ are orthogonal to each other (this is correct both for
$\ket{\mathbf{n}}$ and $\ket{N;l,m}$), we will restrict our discussion to a fixed $N$ first.

It is important to note that states $\ket{N;l,m}$ defined by \eref{eq:SFV-defnlm} are linear
combinations of the $C_3(N)$ states $\ket{\mathbf{n}}$ with $\mathbf{n}^2=N$.  Recall that $C_3(N)$ is the
number of ways to represent the integer $N$ as a sum of squares of three integers.   The states $\ket{N;l,m}$ are not necessarily
orthonormal nor linearly independent.  If we define
\begin{align}
  \mathcal{V}_{N} &= \text{span}\{ \ket{\mathbf{n}}\Big{|}\mathbf{n}^2=N \} \,,\nn\\
  \mathcal{V}_{N;{l_{\text{cut}}}} &= \text{span}\{\ket{N;l,m}\Big{|}l\leq {l_{\text{cut}}},~all~m\} \,,
\end{align}
we will have
\begin{equation}
  \dim(\mathcal{V}_{N;{l_{\text{cut}}}}) \leq \dim(\mathcal{V}_{N}) = C_3(N)~~~~\forall~{l_{\text{cut}}} \in \mathbb{N} \,.
\end{equation}

Moreover, a positive definite Hermitian matrix $P_{N}$ for each $N$ can be introduced to represent their inner products as follows:
\begin{align}\label{eq:SFV-defP}
  [P_{N}]_{l',m';l,m} &= \braket{N;l',m'|N;l,m} \nonumber\\
  &=
  \begin{cases}
    \sum_{\hat{\mathbf{n}}}\,4\pi\, Y_{l'm'}^*(\hat{\mathbf{n}})\, Y_{lm}(\hat{\mathbf{n}}) & N\neq0 \\
    \delta_{l',0}\, \delta_{l,0} & N=0
  \end{cases}
  \,,
\end{align}
$P_{N}$ reflects the degree of partial-wave mixing for a given $N$.  Examples are provided in 
Appendix \ref{sec:CPM}.  For values of $N$ having relatively small values of $C_3(N)$, $[P_{N}]_{l',m';l,m}$
contains many large elements off the diagonal of $l'=l$ and $m'=m$ indicating significant angular
momentum mixing.  However, for other values of $N$, where $C_3(N)$ is relatively large, the
averaging process brings $[P_{N}]_{l',m';l,m}$ closer to the diagonal.
In fact, as $N$ tends to infinity, $P_{N}/C_3(N)$ will approximate an identity matrix with index $(l,m)$.

Considering the definition \eref{eq:SFV-defnlm}, $\ket{N;l,m}$ behaves as the vectors of the irreducible representations (irreps) of $O(3)$.
The irreps of $O(3)$ indicated as spin-parity $J^P$ will decompose into the irreps of $O_h$.
For $l=0,...,4$, the decomposition is
\begin{align}\label{eq:SFV-l2Gamma}
  \mathbf{0}^+ &= \mathbf{A}_1^+ \,, \nn\\
  \mathbf{1}^- &= \mathbf{T}_1^- \,, \nn\\
  \mathbf{2}^+ &= \mathbf{E}^+ \oplus \mathbf{T}_2^+ \,, \nn\\
  \mathbf{3}^- &= \mathbf{A}_2^- \oplus \mathbf{T}_1^- \oplus \mathbf{T}_2^- \,, \nn\\
  \mathbf{4}^+ &= \mathbf{A}_1^+ \oplus \mathbf{E}^+ \oplus \mathbf{T}_1^+ \oplus \mathbf{T}_2^+ \,.
\end{align}
Thus, it is convenient to introduce another basis as follows:
\begin{equation}\label{eq:SFV-defNlG2}
  \ket{l;\Gamma\,,f\,,\alpha} = \sum_m \, [C_l]_{\Gamma,\alpha;m}\ket{l,m} \,,
\end{equation}
with matrix elements of $C_l$ given in \tref{tab:MEC} for $l\leq4$.
Here, $\Gamma$ indicates the irrep, $\alpha$ runs from 1 to the dimension of the irrep $\Gamma$,
and $f$ runs from 1 to the number of occurrences of the irreducible representation $\Gamma$ in the
angular momentum $l$.
For $l\leq 4$, as shown in \tref{tab:MEC}, the value of $f$ always equals 1.
Now, we can define a new basis as follows:
\begin{equation}\label{eq:SFV-defNlG}
  \ket{N,l;\Gamma,f,\alpha} = \sum_m \, [C_l]_{m;\Gamma,f,\alpha}\ket{N;l,m} \,.
\end{equation}
As $C_l$ is a unitary matrix and independent of $N$, we have
\begin{equation}
  \sum_m \ket{N',l;m}\bra{N,l;m} = \sum_{\Gamma,f,\alpha} \ket{N',l;\Gamma,f,\alpha}\bra{N,l;\Gamma,f,\alpha} \,,
\end{equation}
and the potential \eref{eq:SFV-VLlm} will be
\begin{equation}\label{eq:SFV-VLlG}
  \hat{V}_L = \sum_{N',N}\sum_l \,\tilde{v}_l(k_{N'},k_{N}) \sum_{\Gamma,f,\alpha} \ket{N',l;\Gamma,f,\alpha}\bra{N,l;\Gamma,f,\alpha} \, .
\end{equation}
Now $\ket{N,l;\Gamma,f,\alpha}$ with different $N,\Gamma,\alpha$ are orthogonal, so we can define matrix $P_{N;\Gamma,\alpha}$ to represent their inner products,
\begin{align}\label{eq:SFV-defPG}
  [P_{N;\Gamma,\alpha}]_{l',f';l,f} &= \braket{N,l';\Gamma,f',\alpha|N,l;\Gamma,f,\alpha}\,, \nn\\
                                    =& \sum_{m',m}\, [C_{l'}]^*_{m';\Gamma,f',\alpha}\,[P_{N}]_{l',m';l,m}\, [C_l]_{m;\Gamma,f,\alpha}\, . 
\end{align}
That is, we first perform a unitary transformation to the $P$ matrix of \eref{eq:SFV-defP} to
make it block diagonal according to the irreps of the cubic group and then isolate the $\Gamma,\alpha$
part submatrix.

Now we can impose an angular momentum cutoff ${l_{\text{cut}}}$ such that
\begin{equation}
  \hat{V}_L = \sum_{N',N}\sum_{l=0}^{{l_{\text{cut}}}} \,\tilde{v}_l(k_{N'},k_{N}) \sum_{\Gamma,f,\alpha} \ket{N',l;\Gamma,f,\alpha}\bra{N,l;\Gamma,f,\alpha} \,.
\end{equation}
To extract the spectrum in a given representation $\Gamma$, we only need a reduced potential
\begin{align}
  \hat{V}_{L;\Gamma,\alpha} &= \sum_{N',N}\sum_{l=0}^{{l_{\text{cut}}}} \,\tilde{v}_l(k_{N'},k_{N}) \nonumber\\
  &\qquad\times \sum_{f} \ket{N',l;\Gamma,f,\alpha}\bra{N,l;\Gamma,f,\alpha} \,,
\end{align}
for any $\alpha$.

\begin{table}[t]
\caption{Matrix elements of $C_l$ for $l=0,1,2,3,4$. This table is provided in many papers,
  {\it e.g.}, Table A.2 of \cite{Bernard:2008ax}, and the parity is suppressed here.}
\label{tab:MEC}
\centering
\begin{ruledtabular}
\begin{tabular}{cccc}
\noalign{\smallskip}
     $l$ & $\Gamma$ & $\alpha$ & $\sum_m [C_l]_{m;\Gamma,f=1,\alpha}\ket{l,m}$ \\
\noalign{\smallskip}
      \hline
\noalign{\smallskip}
     $0$ & $\mathbf{A}_1$ & $1$ & $\ket{0,0}$ \\
\noalign{\smallskip}
      \hline
\noalign{\smallskip}
     $1$ & $\mathbf{T}_1$ & $1$ & $\frac{1}{\sqrt{2}}(\ket{1,-1}-\ket{1,1})$ \\
     $ $ & & $2$ & $\frac{i}{\sqrt{2}}(\ket{1,-1}+\ket{1,1})$ \\
     $ $ & & $3$ & $\ket{1,0}$ \\
\noalign{\smallskip}
      \hline
\noalign{\smallskip}
     $2$ & $\mathbf{E}$ & $1$ & $\ket{2,0}$ \\
     $ $ & & $2$ & $\frac{1}{\sqrt{2}}(\ket{2,-2}+\ket{2,2})$ \\
     $ $ & $\mathbf{T}_2$ & $1$ & $-\frac{1}{\sqrt{2}}(\ket{2,-1}+\ket{2,1})$ \\
     $ $ & & $2$ & $\frac{i}{\sqrt{2}}(\ket{2,-1}-\ket{2,1})$ \\
     $ $ & & $3$ & $-\frac{1}{\sqrt{2}}(\ket{2,-2}-\ket{2,2})$ \\
\noalign{\smallskip}
      \hline
\noalign{\smallskip}
     $3$ & $\mathbf{A}_2$ & $1$ & $\frac{1}{\sqrt{2}}(\ket{3,-2}-\ket{3,2})$ \\
     $ $ & $\mathbf{T}_1$ & $1$ & $\frac{\sqrt{5}}{4}(\ket{3,-3}-\ket{3,3})-\frac{\sqrt{3}}{4}(\ket{3,-1}-\ket{3,1})$ \\
     $ $ & & $2$ & $\frac{-i\sqrt{5}}{4}(\ket{3,-3}+\ket{3,3})-\frac{i\sqrt{3}}{4}(\ket{3,-1}+\ket{3,1})$ \\
     $ $ & & $3$ & $\ket{3,0}$ \\
     $ $ & $\mathbf{T}_2$ & $1$ & $-\frac{\sqrt{3}}{4}(\ket{3,-3}-\ket{3,3})-\frac{\sqrt{5}}{4}(\ket{3,-1}-\ket{3,1})$ \\
     $ $ & & $2$ & $\frac{-i\sqrt{3}}{4}(\ket{3,-3}+\ket{3,3})+\frac{i\sqrt{5}}{4}(\ket{3,-1}+\ket{3,1})$ \\
     $ $ & & $3$ & $\frac{1}{\sqrt{2}}(\ket{3,-2}+\ket{3,2})$ \\
\noalign{\smallskip}
      \hline
\noalign{\smallskip}
     $4$ & $\mathbf{A}_1$ & $1$ & $\frac{\sqrt{30}}{12}(\ket{4,-4}+\ket{4,4})+\frac{\sqrt{21}}{6}\ket{4,0}$ \\
     $ $ & $\mathbf{E}$ & $1$ & $-\frac{\sqrt{42}}{12}(\ket{4,-4}+\ket{4,4})+\frac{\sqrt{15}}{6}\ket{4,0}$ \\
     $ $ & & $2$ & $-\frac{1}{\sqrt{2}}(\ket{4,-2}+\ket{4,2})$ \\
     $ $ & $\mathbf{T}_1$ & $1$ & $-\frac{1}{4}(\ket{4,-3}+\ket{4,3})-\frac{\sqrt{7}}{4}(\ket{4,-1}+\ket{4,1})$ \\
     $ $ & & $2$ & $\frac{i}{4}(\ket{4,-3}-\ket{4,3})-\frac{i\sqrt{7}}{4}(\ket{4,-1}-\ket{4,1})$ \\
     $ $ & & $3$ & $\frac{1}{\sqrt{2}}(\ket{4,-4}-\ket{4,4})$ \\
     $ $ & $\mathbf{T}_2$ & $1$ & $\frac{\sqrt{7}}{4}(\ket{4,-3}+\ket{4,3})-\frac{1}{4}(\ket{4,-1}+\ket{4,1})$ \\
     $ $ & & $2$ & $\frac{i\sqrt{7}}{4}(\ket{4,-3}-\ket{4,3})+\frac{i}{4}(\ket{4,-1}-\ket{4,1})$ \\
     $ $ & & $3$ & $\frac{1}{\sqrt{2}}(\ket{4,-2}-\ket{4,2})$ \\
\noalign{\smallskip}
\end{tabular}
\end{ruledtabular}
\end{table}

As we know, the angular momentum quantum number $l$ is not a good
quantum number in the finite volume.
Therefore, $l$ cannot appear in a good basis.
Indeed, the states labeled as $\ket{N,l;\Gamma,f,\alpha}$ are not orthogonal.
{\it Shells} defined in \rref{Doring:2018xxx} can show this point clearly.
{\it Shells} can divide states $\ket{\mathbf{n}}$ with $\mathbf{n}^2=N$ further, and there are seven different kinds of {\it shells}: $(0,0,0)\times 1$, $(0,0,c)\times 6$, $(0,b,b)\times 12$, $(0,b,c)\times 24$, $(a,a,a)\times 8$, $(a,a,c)\times 24$, and $(a,b,c)\times 48$, where the numbers indicate how many different states are referred.
As {\it shells} are representations of $O_h$, they can be decomposed into irreps as well. The decomposition is given as follows:
\begin{align}
  (0,0,0) &= \mathbf{A}_1^+ \,,\nn\\
  (0,0,c) &= \mathbf{A}_1^+ \oplus \mathbf{E}^+ \oplus \mathbf{T}_1^- \,,\nn\\
  (0,b,b) &= \mathbf{A}_1^+ \oplus \mathbf{E}^+ \oplus \mathbf{T}_1^- \oplus \mathbf{T}_2^+ \oplus \mathbf{T}_2^- \,,\nn\\
  (0,b,c) &= \mathbf{A}_1^+ \oplus \mathbf{A}_2^+ \oplus 2\mathbf{E}^+ \oplus \mathbf{T}_1^+ \oplus 2\mathbf{T}_1^- \oplus \mathbf{T}_2^+ \oplus 2\mathbf{T}_2^- \,,\nn\\
  (a,a,a) &= \mathbf{A}_1^+ \oplus \mathbf{A}_2^- \oplus \mathbf{T}_1^- \oplus \mathbf{T}_2^+ \,,\nn\\
  (a,a,c) &= \mathbf{A}_1^+ \oplus \mathbf{A}_2^- \oplus \mathbf{E}^+ \oplus \mathbf{E}^- \oplus
  \mathbf{T}_1^+ \oplus 2\mathbf{T}_1^- \nn\\
          &\quad \oplus 2\mathbf{T}_2^+ \oplus \mathbf{T}_2^- \,,\nn\\
  (a,b,c) &= \mathbf{A}_1^+ \oplus \mathbf{A}_1^- \oplus \mathbf{A}_2^+ \oplus \mathbf{A}_2^- \oplus 2\mathbf{E}^+ \oplus 2\mathbf{E}^- \nonumber\\
          &\quad \oplus 3\mathbf{T}_1^+ \oplus 3\mathbf{T}_1^- \oplus 3\mathbf{T}_2^+ \oplus \mathbf{T}_2^- \,.
\end{align}
It is easy to find only one shell in $N=1$, namely, $(0,0,1)$. As there is no $\mathbf{T}_2^+$ in
$(0,0,1)$, the state $\ket{N;l,m} = \ket{1;2,m}$ has no $\mathbf{T}_2^+$, even though $l=2$ has.
Similarly, there is only one $\mathbf{A}_1^+$ in $(0,0,1)$, so $\mathbf{A}_1^+$ in $\ket{1;0,m}$ and
$\ket{1;4,m}$ will be linearly dependent.
Therefore, $\ket{N=1,l=0;\Gamma=\mathbf{A}_1^+,f=1,\alpha=1}$ and
$\ket{N=1,l=4;\Gamma=\mathbf{A}_1^+,f=1,\alpha=1}$ are linearly dependent.

Furthermore, there are two {\it shells} in $N=9$, $(0,0,3)$ and $(2,2,1)$. Some irreps, such as
$\mathbf{A}_1^+$, appear more than once, which means
$\ket{N=9,l=0;\Gamma=\mathbf{A}_1^+,f=1,\alpha=1}$ and
$\ket{N=9,l=4;\Gamma=\mathbf{A}_1^+,f=1,\alpha=1}$ can be linearly independent.
Thus, another state label, $F$, is introduced to replace $l$ and $f$.  The largest value of $F$ is
denoted as $F_{max}$ and this value is governed by $\Gamma$, $l_{cut}$ and $N$, i.e.,
$F_{max}(\Gamma, l_{cut}, N)$.
$F_{max}$ is at most $F_{cut}$, which counts the $l$ and $f$, and does not depend on $N$, i.e.,
$F_{cut}(\Gamma,l_{cut})$.

For example, when $\Gamma=\mathbf{A}_1^+$ and $l_{cut}=4$, $F_{cut}(\mathbf{A}_1^+,4)=2$, as
$\mathbf{A}_1^+$ appears twice in \eref{eq:SFV-l2Gamma}.  For
$N=9$, $F_{max}(\mathbf{A}_1^+,4,9)=2$, as there are two independent states originating from {\it shells}
$(0,0,3)$ and $(2,2,1)$.  However, $F_{max}(\mathbf{A}_1^+,4,1)=1$ for $N=1$.
In Appendix \ref{sec:app}, we give a detailed calculation for $N=1$ and $9$ cases.

Now, we proceed to orthonormalize the basis $\ket{N,l;\Gamma,f,\alpha}$ to get the final basis
labeled as $\ket{N;\Gamma,F,\alpha}$.  From the Wigner-Eckart theorem,
$v_{\Gamma,F',F}(k_{N'},k_{N})$ is $\alpha$ independent and we have
\begin{align}
  \hat{V}_L &= \sum_{N',N} \sum_{\Gamma,F',F} \,v_{\Gamma,F',F}(k_{N'},k_{N}) \nonumber\\
            &\qquad\times \sum_\alpha \ket{N';\Gamma,F',\alpha}\bra{N;\Gamma,F,\alpha} \,.
\end{align}
Using \eref{eq:SFV-VLlG}, 
\begin{align}\label{eq:SFV-defV}
  v_{\Gamma,F',F}(k_{N'},k_{N}) &= \bra{N';\Gamma,F',\alpha}\hat{V}_L\ket{N;\Gamma,F,\alpha} \nn\\
                                &=\sum_{l=0}^{{l_{\text{cut}}}} \,\tilde{v}_l(k_{N'},k_{N}) \nonumber\\
                                &\qquad\times \sum_{f} \braket{N';\Gamma,F',\alpha|N';l,\Gamma,f,\alpha} \nonumber\\
                                &\qquad\qquad\times \braket{N;l,\Gamma,f,\alpha|N;\Gamma,F,\alpha}\,, \nn\\
                                &=\sum_{l=0}^{{l_{\text{cut}}}} \,\tilde{v}_l(k_{N'},k_{N})\,  [G_{l;\Gamma}]_{N',F';N,F}~~~~\forall \alpha \,,
\end{align}
where
\begin{equation}\label{eq:SFV-defGB}
  [G_{l;\Gamma}]_{N',F';N,F} = \sum_{f}\, [M_{l;\Gamma,\alpha}]^*_{f;N',F'}\, [M_{l;\Gamma,\alpha}]_{f;N,F}~~~~\forall \alpha\,,
\end{equation}
with
\begin{equation}\label{eq:SFV-defGS}
  [M_{l;\Gamma,\alpha}]_{f;N,F} = \braket{N,l;\Gamma,f,\alpha|N;\Gamma,F,\alpha}\,.
\end{equation}
The inner product matrix \eref{eq:SFV-defPG} tells us not only how to do the orthonormalization,
but also how to compute the mixed inner product in \eref{eq:SFV-defGS}, so it summarizes all the
things needed to solve for $G$.

As a specific implementation of the orthonormalization procedure, we present an eigenmode-based
method here.  An alternative approach based on the Gram-Schmidt procedure is presented in 
Appendix \ref{sec:app}.  We discuss the case $F_\text{cut}(\Gamma,{l_\text{cut}})=1+1$ (we use
$1+1$ to mean that the two $l$ containing $\Gamma$ are different) here, and the generalization
should be straightforward.  One proceeds by selecting particular values for $N$, $\Gamma$, $\alpha$,
and $f$ and constructing the inner-product matrix
\begin{equation}\label{eq:rPN}
  \tilde{P}_{N;\Gamma,\alpha} =
  \begin{pmatrix}
    \braket{l_1|l_1} & \braket{l_1|l_2} \\
    \braket{l_2|l_1} & \braket{l_2|l_2}
  \end{pmatrix}\,.
\end{equation}
Here indices $N$, $\Gamma$, $\alpha$ and $f$ have been suppressed in the bra-ket notation, {\it
  i.e.} $\ket{l_1} = \ket{N, l_1; \Gamma, f, \alpha}$.
One then solves the eigenvalue equation
\begin{equation}
\tilde{P}_{N;\Gamma,\alpha}\,X^i = \lambda^i\,X^i\,,
\end{equation}
providing the orthonormalized eigenvectors $X^i$ with eigenvalues $\lambda^i$.  States $\ket{N;
  \Gamma, F , \alpha} \equiv \ket{F}$ can be constructed from
\begin{eqnarray}
  \ket{\tilde{F}_1} &=& X^1_j\,\ket{l_j}\,, \nn \\
  \ket{\tilde{F}_2} &=& X^2_j\,\ket{l_j}\,.
\end{eqnarray}
These states are easily normalized via the consideration of
\begin{eqnarray}
  \braket{\tilde{F}_i|\tilde{F}_i} &=& X_j^{i*}\,\braket{l_j|l_k}\,X_k^i\,, \nn \\
                       &=& X_j^{i*}\,\left[\tilde{P}_{N;\Gamma,\alpha}\right]_{jk}\,X_k^i\,, \nn \\
                       &=& \lambda^i\, X^{i\dagger}\, X^i
                        =  \lambda^i\,.
\end{eqnarray}
Thus, the orthonormal vectors $\ket{N; \Gamma, F_i, \alpha}$ are
\begin{equation}
  \ket{N; \Gamma, F_i, \alpha} = \left\{
    \begin{array}{lr}
      0\,, & \lambda^i = 0 \\
      \frac{X_j^i}{\sqrt{\lambda^i}}\,\ket{N, l_j; \Gamma, f, \alpha}\,, & \lambda^i \neq 0
    \end{array} \right. \,.
\end{equation}
$F_{\rm max}$ is given by the number of nonzero eigenvalues, {\it i.e.} the rank of $\tilde{P}_{N;\Gamma,\alpha}\,.$

Now we have the correct orthonormal basis $\ket{N;\Gamma,F,\alpha}$ 
with the mixed inner products of \eref{eq:SFV-defGS} given by
\begin{eqnarray}
	[M_{l_j;\Gamma,\alpha}]_{f;N,F_i} 
		&=&\braket{N,l_j;\Gamma,f,\alpha|N;\Gamma,F_i,\alpha} \nn\\
       &=& \sum_k \frac{1}{\sqrt{\lambda^i}}\, \left[\tilde{P}_{N;\Gamma\alpha}\right]_{jk}\, X_k^i\,, \nn\\
       &=& \frac{\lambda^i\, X^i_j}{\sqrt{\lambda^i}}
       = \sqrt{\lambda^i}\,X^i_j\,.
\end{eqnarray}

With a momentum cutoff $N_{\text{cut}}$ imposed such that $N\leq N_{\text{cut}}$, a
Hamiltonian of dimension at most $N_{\text{cut}}~F_{\text{cut}}(\Gamma,{l_{\text{cut}}})+1$ is
generated to extract the spectrum of the representation $\Gamma$.  Here,
$F_{\text{cut}}(\Gamma,{l_{\text{cut}}})$ counts the representation $\Gamma$ in all
$l\leq{l_{\text{cut}}}$ and the $+ 1$ accounts for $N=0$.

In summary, the general approach proceeds as follows. First, one performs the summation in
\eref{eq:SFV-defP} to get the $(l_{\text{cut}}+1)^2\times(l_{\text{cut}}+1)^2$ matrices $P_{N\leq
  N_{\text{cut}}}$.  Second, the unitary transformation of \eref{eq:SFV-defPG} is performed to make
these matrices block diagonal according to the irreps of the cubic group.  One then considers the
$\Gamma,\alpha$ portions, which are $F_{\text{cut}}\times F_{\text{cut}}$ matrices.  Finally, one
uses these inner product matrices to orthonormalize the states $\ket{N,l;\Gamma,f,\alpha}$ to
construct the final $\ket{N;\Gamma,F,\alpha}$ basis states and compute the combination
coefficients $G$ through Eqs. \ref{eq:SFV-defGB} and \ref{eq:SFV-defGS}. 

At last, we generalize our discussion for particles with spin.
The first step is to add the spin quantum number $(s,s_z)$ to $\ket{\mathbf{k}}$ to give $\ket{\mathbf{k};s,s_z}$, then one combines them with spherical harmonics as in \cref{eq:IFVSP} to define 
\begin{align}\label{eq:nsb}
  \ket{k;l,l_z;s,s_z} := \int d\Omega_{\hat{\mathbf{k}}}\, Y_{l,l_z}(\hat{\mathbf{k}})\ket{\mathbf{k};s,s_z} \,,
\end{align}
and it can be further combined with the Clebsch–Gordan coefficients $\braket{l,l_z;s,s_z|j,j_z}$ to produce
\begin{align}
  \ket{k;l,s;j,j_z} := \sum_{l_z,s_z} \ket{k;l,l_z;s,s_z} \braket{l,l_z;s,s_z|j,j_z} \,.
\end{align}
Since $(j,j_z)$ are now the good rotation quantum numbers, Wigner-Eckart theorem will only allow the interactions built with
\begin{align}
  \ket{k';l',s';j,j_z}\bra{k;l,s;j,j_z} \,.
\end{align}
We also note that when in the finite volume, one will need the $P$-matrix for $\ket{N;l,s;j,j_z}$ (that is the finite-volume counterpart for $\ket{k;l,s;j,j_z}$) defined as
\begin{align}
  [P_{N;s}]_{l',j',j_z';l,j,j_z} &= \braket{N;l',s;j',j_z'|N;l,s;j,j_z} \nonumber\\
  &= \sum_{l_z',l_z;s_z} \braket{j',j_z'|l',l_z';s,s_z} [P_N]_{l',l_z';l,l_z} \nonumber\\
  &\quad \times \braket{l,l_z;s,s_z|j,j_z} \,,
\end{align}
where $P_N$ is the $P$-matrix defined in \cref{eq:SFV-defP}.
Second, one constructs the cubic basis as in \cref{eq:SFV-defNlG} to be
\begin{align}\label{eq:nse}
  \ket{N;l,s;\Gamma,f,\alpha} = \sum_m \, [C_j]_{\Gamma,f,\alpha;j_z}\ket{N;l,s;j,j_z} \,,
\end{align}
where the coefficient matrix $C_j$ can be found in many papers, {\it e.g.} Table A.2 (for bosons) and Table A.4 (for fermions) of Ref.~\cite{Bernard:2008ax}.
It is now straightforward to get the $P$-matrix for $\ket{N;l,s;\Gamma,f,\alpha}$, to use it to orthonormalize these states, and to obtain the final combination coefficients.

\section{Example of Isospin-2 {\large $\mathbf{\pi\pi}$} Scattering}\label{sec:EIS}

In this section, the formalism developed herein is applied to analyze lattice QCD results for the
isospin-2 $\pi\pi$ scattering system.
In doing so we will explore the consistency of a separable potential analysis result with that from L\"{u}scher's method.

\begin{figure}[t]
  \centering
  \includegraphics[width=\columnwidth]{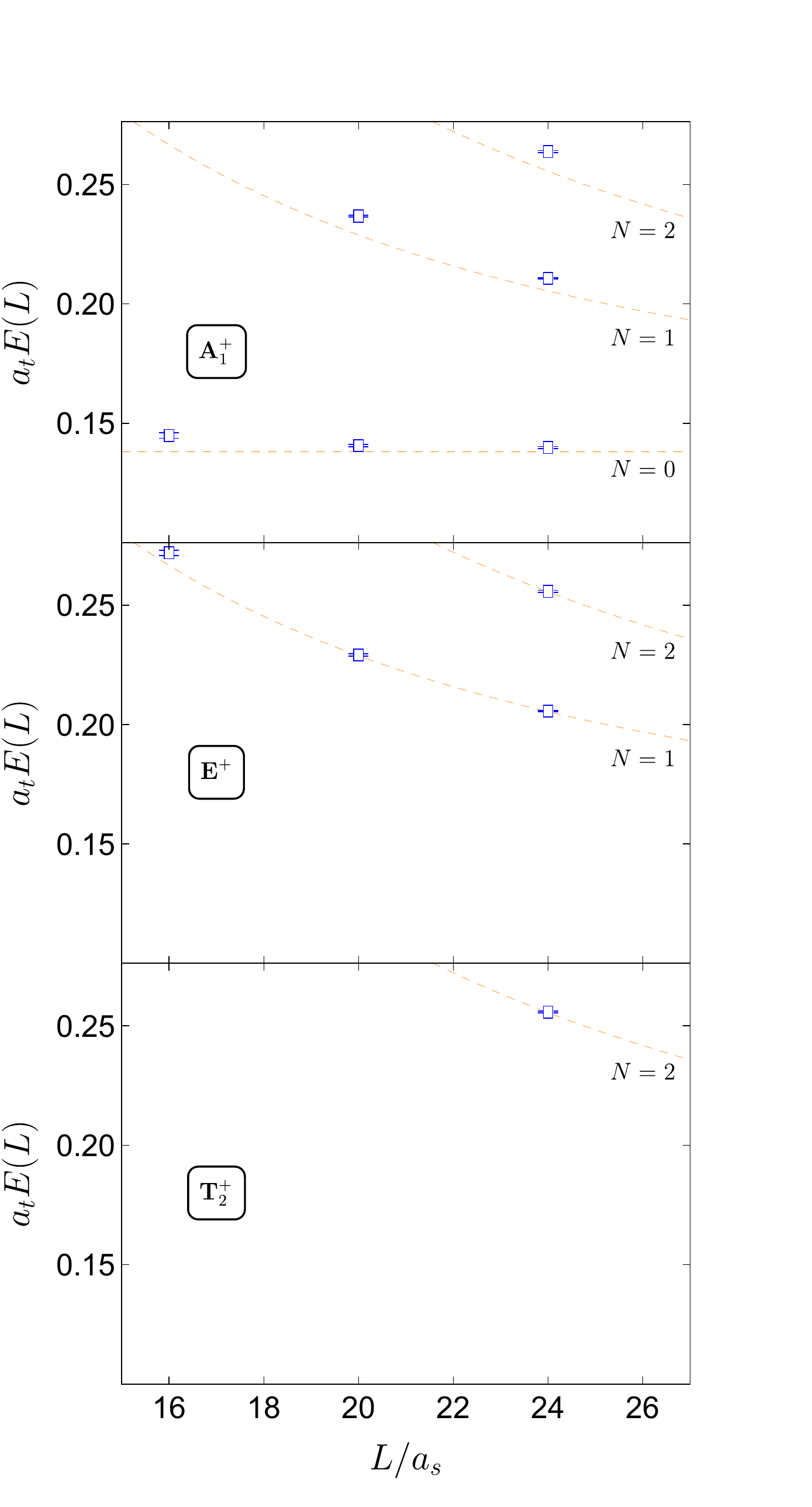}
  \caption{Lattice spectrum for irreps $\mathbf{A}_1^+$, $\mathbf{E}^+$, $\mathbf{T}_2^+$ from Ref.~\cite{Dudek:2012gj}. Dashed
    curves represent the noninteracting rest-frame pion-pair energies $2\sqrt{m_\pi^2+k_N^2}$.}
  \label{fig:eDudek}
\end{figure}

The lattice QCD results are from \refref{Dudek:2012gj} where an anisotropic action is used.  They
quote the spatial lattice spacing $a_s\sim0.12~$fm, temporal lattice spacing $a_t^{-1}\sim5.6~$GeV,
and the anisotropy $\xi={a_s}/{a_t}=3.444(6)$.  When setting the scale, we refer to $a_t$ and
$\xi$.
The pion mass for the simulation results $a_t\, m_\pi = 0.06906(13) \sim 396$~MeV.
In our analysis, only $\xi=3.444(6)$ and $a_t\, m_\pi = 0.06906(13)$ will be used, since we did not find sufficiently precise values for $a_s$ and $a_t$ in Ref.~\cite{Dudek:2012gj}.
In the analysis of \refref{Dudek:2012gj} lattice results above the $4\pi$ threshold are not
included and since our formalism does not include the four-body contributions, we apply the
same cut.
%
%
The results we fit are illustrated in \fref{fig:eDudek}.

\subsection{Separable potential analysis}

Following \refref{Dudek:2012gj}, we work with dimensionless lattice units. The energy $h$ of
\eref{Eq:IFV} is taken as
\begin{equation}
  a_t \,h(k) = 2\sqrt{(a_t\, m_\pi)^2+\left( a_t\, k \right)^2}\,,
\end{equation}
and when going to the finite-volume system, we have
\begin{equation}
  a_t\, k \to a_t\, k_N = a_t\, \frac{2\pi\, \sqrt{N}}{L}\, .
\end{equation}
Only $s$, $d$, and $g$ waves will be taken into account as in \refref{Dudek:2012gj}. With the
partial-wave expansion of \eref{eq:SFV-PW}, the partial-wave potentials are taken to be of a simple
separable form
\begin{equation}
  a_t^{-2}\, v_l(p,k) = \frac{G_l}{(a_t\, m_\pi)^2}\, f_l(p)\, f_l(k) \,,
\end{equation}
with
\begin{align}
  f_0(k) = \frac{1}{(1+(d_0\, a_t\, k)^2)^2} \,, \nn\\
  f_2(k) = \frac{(d_2\, a_t\, k)^2}{(1+(d_2\, a_t\, k)^2)^3} \,, \nn\\
  f_4(k) = \frac{(d_4\, a_t\, k)^4}{(1+(d_4\, a_t\, k)^2)^4} \,,
\end{align}
with parameters $G_l$ and $d_l$ dimensionless.

\begin{figure}[t]
  \centering
  \includegraphics[width=\columnwidth]{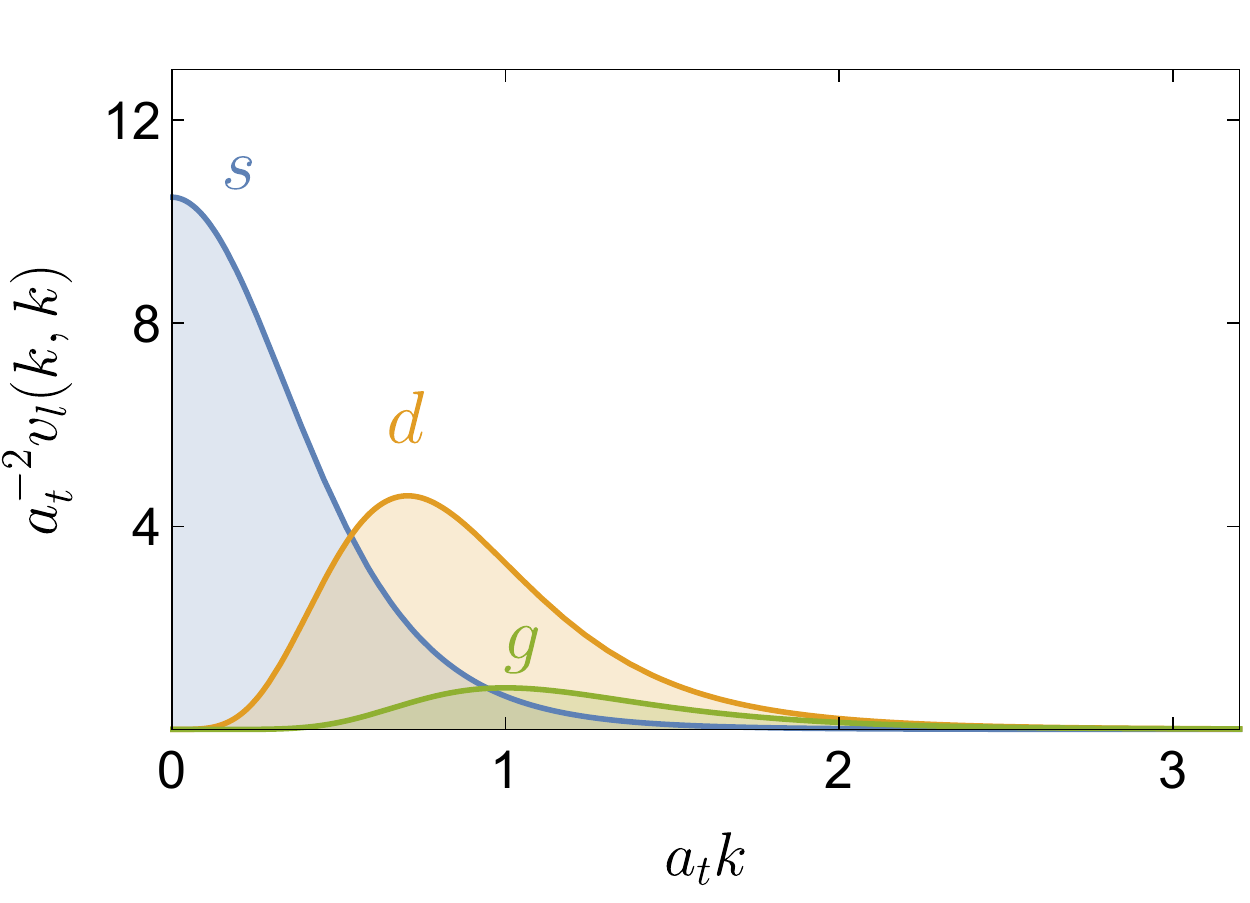}
  \caption{The $s$, $d$, and $g$ partial-wave potential, $a_t^{-2} v_l(k,k)$, with illustrative
    parameters $G_0=1/20$ and $G_2=G_4=d_0=d_2=d_4=1$.}
  \label{fig:potential}
\end{figure}

There are two fit parameters $G_l$ and $d_l$ for each partial wave.  The amplitude of the potential
is governed by $G_l$.
The $s$-wave potential reaches its peak value at $k=0$, while the peak positions of the $d$- and
$g$-wave potentials are determined by $d_l$.
The shapes of these potentials are shown in \fref{fig:potential}.  for $G_0 \sim G_2 \sim G_4$ the
potentials have the natural progression $v_0>v_2>v_4$.

The spectrum is solved from the reduced Hamiltonian discussed in \sref{sec:MPW}.
Then the phase shifts are determined through the Lippmann-Schwinger as discussed around \eref{eq:LSE}.
With the cutoff $N_{\text{cut}}$ chosen to be $600$ ($\sim 10$ GeV) as in \refref{Wu:2014vma}, the
full Hamiltonian involves $\sum_{N=0}^{600} C_3(N) = 61565$ states and can be reduced to
Hamiltonians of dimension $923$, $965$, and $963$ for $\mathbf{A}_1^+$, $\mathbf{E}^+$, and
$\mathbf{T}_2^+$ respectively.

\subsection{Fitting to Lattice QCD Results}

Our fitting procedure is to minimise the $\chi^2$ defined by
\begin{equation}
  \chi^2 = [E_{\text{Sep}}-E_{\text{Lattice}}]^T\, [\mathbb{C}]^{-1}\, [E_{\text{Sep}}-E_{\text{Lattice}}] \,,
\end{equation}
where $E_{\text{Sep}}-E_{\text{Lattice}}$ denotes the vector of the differences between the
spectrum obtained in the separable potential model and the lattice.  The covariance matrix $\mathbb{C}$ denotes the
covariances in the lattice spectrum of \refref{Dudek:2012gj}.

When fitting with all six parameters free, the phase shifts $\delta_2$ and $\delta_4$ show an
unreasonable behaviour in the high-momentum range due to the exclusion of high-energy lattice QCD
results above the $4 \pi$ threshold which would otherwise constrain the regulator parameters $d_2$
and $d_4$.
In the absence of lattice constraints we consider three models for the parameters $d_2$ and $d_4$.

Introducing the more familiar regulator parameter $\Lambda$ via
\begin{equation}
\frac{k}{\Lambda_i} = d_i\, a_t k \, ,
\end{equation} 
we consider values $d_A=7.17,~d_B=4.78$, and $d_C=3.58$ corresponding to 
$\Lambda_A \sim 0.8 \text{ GeV},\ \Lambda_B \sim 1.2 \text{ GeV}$, and $\Lambda_C \sim 1.6 \text{
  GeV}$, respectively.  

\begin{figure}[t]
  \centering
  \includegraphics[width=\columnwidth]{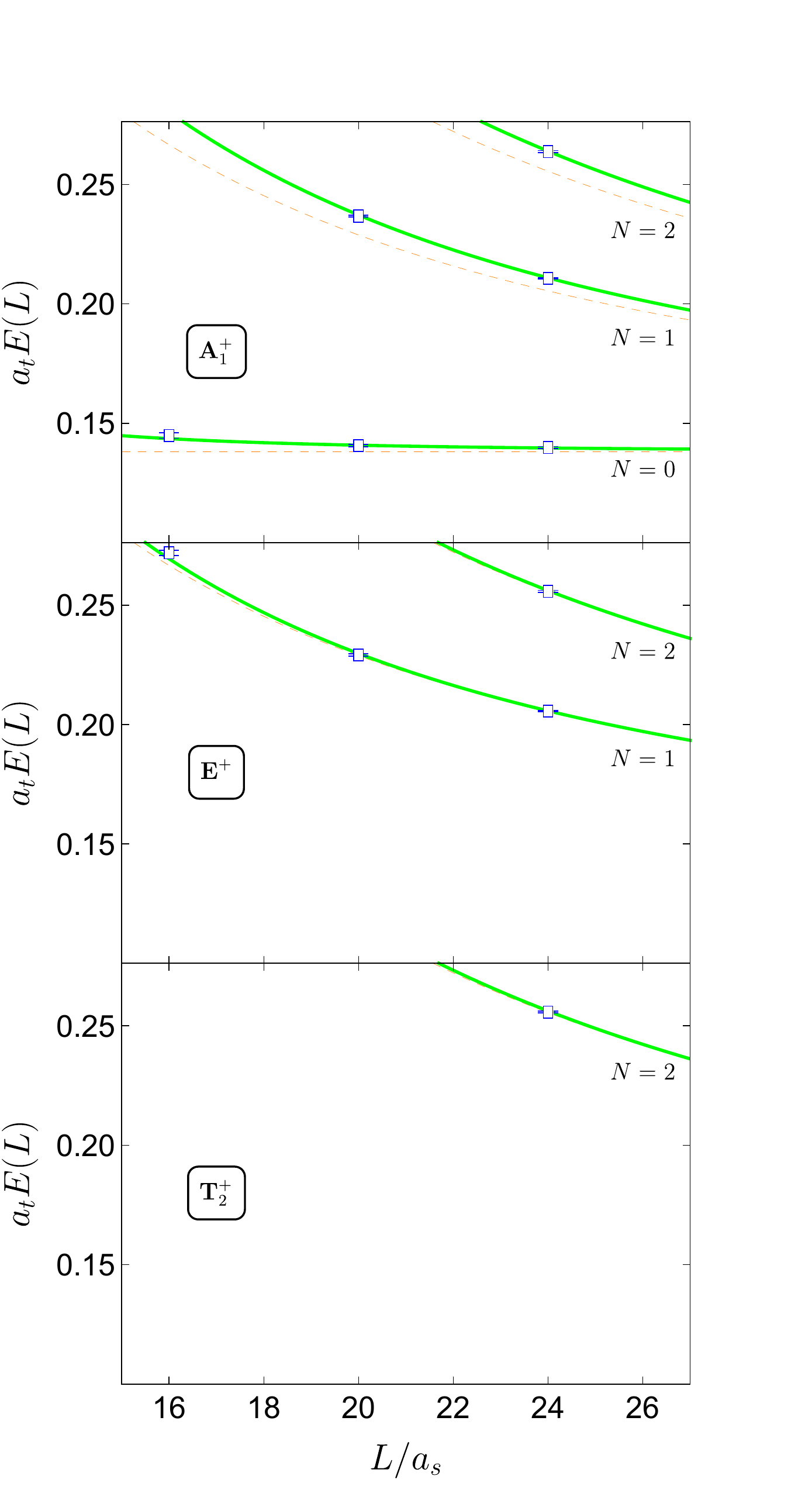}
  \caption{Finite-volume spectrum fit of the separable potential model to the lattice QCD results of \refref{Dudek:2012gj}
    for isospin-2 $\pi\pi$ scattering.  Solid curves illustrate the energies resolved in the separable potential model as
    the fit parameters of Table \ref{tab:fitres} Case $B$ are optimised to fit the lattice QCD
    results (square points).  Dashed curves illustrate the non-interacting rest-frame pion-pair
    energies as in \fref{fig:eDudek}.}
  \label{fig:eFit}
\end{figure}

\begin{figure*}[t]
  \centering
  \includegraphics[width=\textwidth]{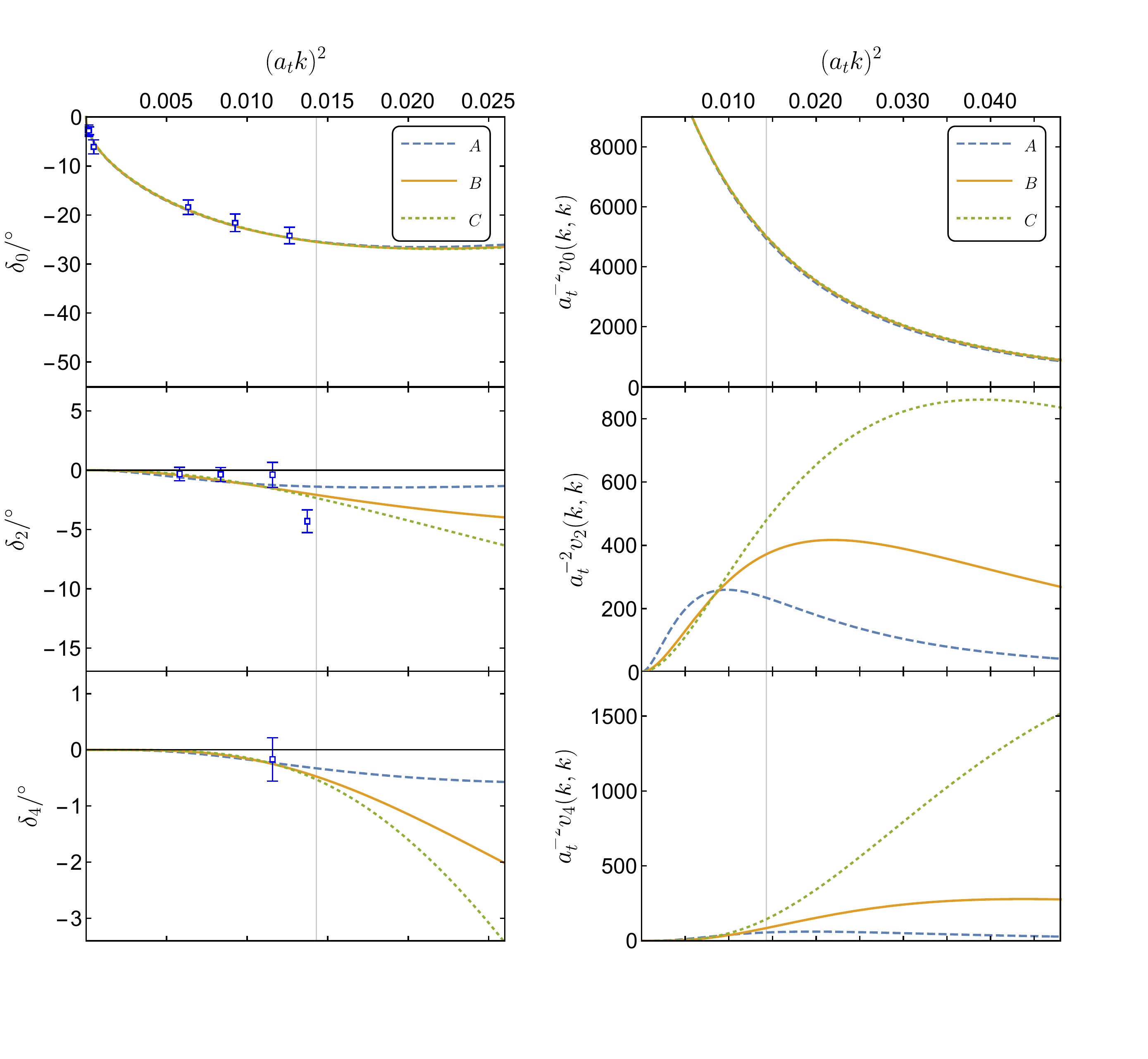}
  \caption{Phase shifts (left) and potentials (right) for $s$, $d$, and $g$ waves from the separable potential model are
    illustrated by the curves for the three cases considered as described in the text.  The square
    points from \refref{Dudek:2012gj} are those that can be extracted from the finite-volume
    spectrum of \fref{fig:eDudek} using L\"{u}scher's method.
    The vertical lines denote the $4\pi$ threshold with $(a_t\,k)^2=0.0143$.}
\label{fig:fixed}
\end{figure*}

We proceed with $d_0$ as a fit parameter and constrain $d_2 = d_4 = d_i$ with $i=A,\ B$, or $C$.
The resulting parameters are shown in \tref{tab:fitres} 
The volume-dependent spectra are shown in \fref{fig:eFit} using the parameters for case $B$ with
$\Lambda_B \sim 1.2$ GeV.
The phase shifts and potentials are illustrated in \fref{fig:fixed} for all three cases considered.

\begin{table}[t]
  \caption{Parameters optimising the fit of the separable potential model to the lattice QCD results of
    \refref{Dudek:2012gj} for isospin-2 $\pi\pi$ scattering.  Three cases for the regular
    parameters $d_2$ and $d_4$ are considered as described in the text.  In each case, four
    parameters are constrained by 11 lattice QCD results leaving seven degrees of freedom.}
\label{tab:fitres}
\centering
\begin{ruledtabular}
\begin{tabular}{cccccccc}
     &          &\multicolumn{2}{c}{$\ell = 0$} &\multicolumn{2}{c}{$\ell = 2$} &\multicolumn{2}{c}{$\ell = 4$} \\
Case & $\chi^2$ & $G_0$ & $d_0$ & $G_2$ & $d_2$ & $G_4$ & $d_4$ \\
\noalign{\smallskip}
\hline
\noalign{\smallskip}
$A$ & 13.4 & 68.5 & 4.63 & 56.4 & $d_A$ & $7.43\times 10^1$ & $d_A$\\
$B$ & 10.5 & 67.8 & 4.57 & 90.6 & $d_B$ & $3.40\times 10^2$ & $d_B$\\
$C$ &  9.8 & 67.2 & 4.54 & 187. & $d_C$ & $2.34\times 10^3$ & $d_C$ \\
\end{tabular}
\end{ruledtabular}
\end{table}

In the top pair of figures in \fref{fig:fixed}, the phase shifts and potentials for $s$-wave scattering are given.
Because of the similar fit parameters in the three cases considered, it is not surprising that
their corresponding phase shifts and potentials are similar. 
It illustrated how these lattice QCD results constrain the $s$-wave phase shift of $\pi\pi$
scattering well.  Moreover the separable potential model result is consistent with that of Ref.~\cite{Dudek:2012gj}
employing L\"{u}scher's method.

In the remaining panels of \fref{fig:fixed}, the $d$- and $g$-wave phase shifts and potentials are shown.
They show different features from the $s$-wave case.  In the low-momentum range where lattice QCD
results constrain the effective field theory, all three cases predict phase shifts in a consistent
manner and in agreement with L\"{u}scher's method for the phase shifts away from the $4\pi$ cut
\cite{Dudek:2012gj}. Beyond the $4\pi$ cut, the various cases diverge.

The behaviour of the potentials provides an explanation for this.  The lattice QCD results
prefer a potential increasing steadily in the low-momentum range.  However, there is freedom to
lower the value of $d$ with a suitable increase in $G$ at the same time to maintain the behaviour
of the potential in the low momentum regime.
If we increase $d$, we do get smaller $\chi^2$, but that may not be reasonable.
%
If one expects $v_0>v_2>v_4$, $d_2 = d_4 = d_B$ is favoured over $d_C$.
To constrain these parameters, one needs information from the lattice at higher energies.

\begin{figure}[t]
\centering
\includegraphics[width=\columnwidth]{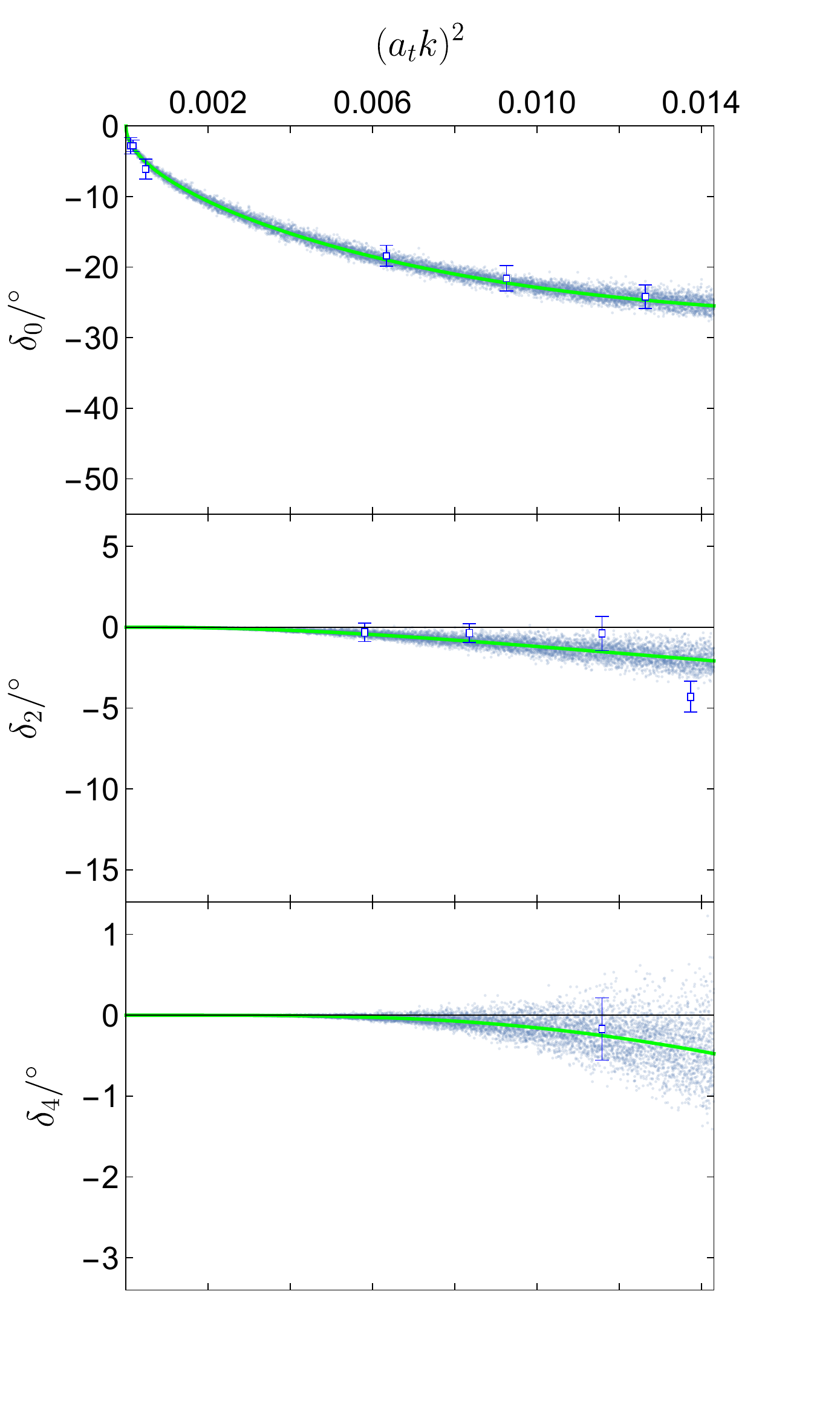}
\caption{Phase shifts predicted by the separable potential model for $s$ (top), $d$ (middle) and $g$ (bottom) partial waves.
  Solid curves illustrate the central value of our fit as in \fref{fig:fixed} and the scattered
  points describe the uncertainty.}
\label{fig:deltaError}
\end{figure}

We now proceed to estimate the uncertainties associated with the separable potential analysis.  Here we focus on
the preferred case with $\Lambda_B \sim 1.2$ GeV.

Our covariance for parameters \( \lambda_i \) is defined as $[\mathcal{H}/2]^{-1}$, where
$[\mathcal{H}]$ is the Hessian of $\chi^2$, the matrix of second-order partial derivatives over
parameters
\begin{equation}
[\mathcal{H}]_{i,j}=\frac{\partial^2\chi^2}{\partial \lambda_i\, \partial \lambda_j} \,.
\end{equation}
As two $d_l$ are fixed, we only have four parameters and the final covariance returned by MINUIT 2
(ordered as $G_0$, $d_0$, $G_2$, $G_4$) is
\begin{equation}\label{eq:Hessian}
\left(
\begin{array}{cccc}
\underline{11.4} & \underline{0.674} & 58.4 & -1.97*10^2 \\
\underline{0.674} & \underline{0.0773} & 1.53 & -9.61 \\
58.4 & 1.53 & \underline{8.02*10^2} & -1.28*10^3 \\
-1.97*10^2 & -9.61 & -1.28*10^3 & \underline{9.45*10^4} \\
\end{array}
\right) \,.
\end{equation}
As the different values of $l$ are decoupling in solving for the phase shifts, the values
underlined in \eref{eq:Hessian} are used in calculating the errors in the phase shifts.  To obtain
error estimates, we re-sample the parameters by using \eref{eq:Hessian} to give $10000$ sets of
parameters for each $l$.  These parameter sets are then used to solve for the phase shifts at
$10000$ different momenta, as illustrated by the scattered points in \fref{fig:deltaError}.
Results from our analysis and from Ref.~\cite{Dudek:2012gj} are consistent, as shown in \cref{fig:deltaError}. Thus the reduction formalism is working well and the Hamiltonian formalism is indeed consistent with the L\"{u}scher method.
Furthermore, one advantage of our method might convey is that to formally disentangle unphysical partial wave mixing using the L\"{u}scher method without any assumptions (even EFT assumptions), one needs different physical volumes where the same scattering momentum is allowed. Arranging this for each energy level is, computationally, impractical, at least in a Lattice QCD setting.
On the other hand, it is worth mentioning that the $\pi\pi$ phase shifts are all very small in our calculation.
Noting our energy range is below the 4$\pi$ threshold, we expect Chiral perturbation theory ($\chi$PT) to be applicable in this energy regime. Thus it is no surprise that the spectrum and resulting phase shifts can be described by our phenomenological model with four free parameters. 

We can also examine the free state constituents of the energy eigenstates. \tref{tab:states}
describes the composition of the first three levels for $L=24$ and $\Gamma=\mathbf{A}_1^+$.  Note
that the further an eigenenergy is away from the free energy, the more complex is its structure.  Only
when an eigenstate is close in energy to the non-interacting basis state will the eigenstate be
dominated by single free state.  For the first few energy eigenstates in the other representations
considered, it is not surprising that they are almost $100\%$ composed by a single free state as
their energies are almost the free energies
.
\begin{table}[t]
\caption{The non-interacting (free) basis-state composition of the first three energy eigenstates 
for $L/a_s=24$ and $\Gamma=\mathbf{A}_1^+$. Basis states with momenta $k_N$ span the columns.
Each row describes the contributions of the basis states as a percentage to an energy eigenstate.}
\label{tab:states}
\centering
\begin{ruledtabular}
\begin{tabular}{ccccccc}

Eigenstate & $N=0$ & $N=1$ & $N=2$ & $N=3$ & $N=4$ & $\cdots$\\
\noalign{\smallskip}
\hline
\noalign{\smallskip}
{1st} & 99.7 & 0.2  & 0.0  & 0.0 & 0.0 & $\cdots$\\
{2nd} & 0.1  & 97.4 & 1.9  & 0.2 & 0.0 & $\cdots$\\
{3rd} & 0.0  & 1.5  & 94.5 & 2.8 & 0.3 & $\cdots$\\
\end{tabular}
\end{ruledtabular}
\end{table}

\section{Summary}\label{sec:Sum}

In this work, a formalism based on Hamiltonian Effective Field Theory has been developed to address
partial-wave mixing in a periodic finite volume.  The formalism is required to connect
infinite-volume partial-wave scattering phase shifts to the finite-volume spectra of lattice QCD.
Our formalism has been developed with reference to the rest frame.

A key step is to introduce a momentum cut and an angular-momentum cut in the finite volume, 
enabling one to reduce the dimension of the Hamiltonian in the finite volume.
Indeed, the formalism presented herein provides an optimal set of rest-frame basis states maximally
reducing the dimension of the Hamiltonian.
To obtain the optimal rest-frame basis states, we introduced several intermediate bases as shown in
in \eref{eq:SFV-defnlm} and \eref{eq:SFV-defNlG}.  Calculation of their inner products not only
illustrates partial wave mixing but also enables the determination of the link from the original
infinite-volume Hamiltonian, where the scattering potentials are defined, to the finite-volume of the
lattice where irreps of the cubic group are considered.
Finally, we presented a specific step-by-step implementation of the process to determine the
infinite-volume potentials contributing to the irreps of the finite-volume.

We then considered an in-practice example of isospin-2 $\pi\pi$ scattering to test the consistency
between a separable potential model and L\"{u}scher's method as implemented in \refref{Dudek:2012gj}.
The results demonstrate that the formalism developed herein  is consistent 
with that of L\"{u}scher.


While relating the finite-volume spectrum and phase shifts as in L\"{u}scher's method, a Hamiltonian formalism also
provides insight into the composition of the finite-volume energy eigenstates.  In this analysis,
one observes only mild mixing between the non-interacting basis states. This provides a deeper
understanding of why current lattice methods for exciting these two-particle states are so
effective.

Additional information is available from the consideration of scattering systems in moving frames.
In light of the limited information available in contemporary lattice QCD simulations in the rest
frame, we consider the development of the moving frame formalism to be necessary.
The moving frame formalism is also necessary for a three-body formalism, since two of the three particles can have a nonvanishing total momentum. In the three-body case, a direct Hamiltonian fit should be formally simpler than the three-body L\"{u}scher formalism. Of course, one of the challenges is the significant increase of the dimension of the Hamiltonian matrix.

\section*{ACKNOWLEDGEMENTS}
It is a pleasure to thank Stephen Sharpe for interesting discussions on the research presented
herein during his visit as a George Southgate Fellow.  The finite-volume energy levels and their
covariances from \refref{Dudek:2012gj} were provided by the Hadron Spectrum Collaboration -- no
endorsement on their part of the analysis presented in the current paper should be assumed.
%
%
This
research was supported by the Australian Research Council through ARC Discovery Project Grants
Nos.\ DP150103101 and DP180100497 (A.W.T.) and DP150103164 and DP190102215 (D.B.L.).
This project was also supported by the Australian Government Research Training Program Scholarship.
\begin{appendix}

\section{Example of Solving For the Combination Coefficients $\mathbf{G}$ through the Gram-Schmidt Process}
\label{sec:app}

We will discuss $F_\text{cut}(\Gamma,{l_\text{cut}})=1$ and
$F_\text{cut}(\Gamma,{l_\text{cut}})=1+1$ (we use $1+1$ to mean that the two $l$ containing
$\Gamma$ are different) cases in detail, these cases cover all cases of ${l_\text{cut}}\leq4$.
The P-Matrix \eref{eq:SFV-defPG} is assumed to be solved already, as the calculation process is
straightforward by using \eref{eq:SFV-defP}, \eref{eq:SFV-defPG} and \tref{tab:MEC}.


\subsection{Case of $\mathbf{F_\text{cut}(\Gamma,{l_\text{cut}})=1}$}

	There is only one $l\leq{l_\text{cut}}$ containing $\Gamma$ now.
	For any $\alpha$, $\ket{N;\Gamma,F=1,\alpha}$ are just normalised $\ket{N;l,\Gamma,f=1,\alpha}$. So from \eref{eq:SFV-defGS}, we have
	\begin{equation}
		[G_{l;\Gamma,\alpha}]_{f=1;N,F=1} = \sqrt{[P_{N;\Gamma,\alpha}]_{l,f=1;l,f=1}} \,,
	\end{equation}
	and states with $[G_{l;\Gamma,\alpha}]_{f=1;N,F=1}=0$ should be discarded.
	Then
	\begin{align}
		&[G_{l;\Gamma}]_{N',F'=1;N,F=1} \nn\\ = &\sqrt{[P_{N';\Gamma,\alpha}]_{l,f=1;l,f=1}\, [P_{N;\Gamma,\alpha}]_{l,f=1;l,f=1}} \,,
	\end{align}
	so we have
	\begin{align}
          \hat{V}_{L;\Gamma,\alpha} &= \sum_{N',N}\, \tilde{v}_l(k_{N'},k_{N})\, \nonumber\\
                                    &\qquad\times \sqrt{[P_{N';\Gamma,\alpha}]_{l,f=1;l,f=1}\, [P_{N;\Gamma,\alpha}]_{l,f=1;l,f=1}} \nonumber\\
                                    &\qquad\qquad\times \ket{N';\Gamma,f',\alpha}\bra{N;\Gamma,f,\alpha} \,.
	\end{align}

	\subsubsection*{Cases of Pure $s$-wave and Pure $p$-wave}\label{sec:CPP}

        \noindent
	As
	\begin{align}
          [P_{N;\mathbf{A}_1^+,\alpha=1}]_{l=0,f=1;l=0,f=1} &= [P_{N}]_{l=0,m=0;l=0,m=0} \nonumber\\
          &= C_3(N) \,,
	\end{align}
	and
	\begin{align}
          [P_{N;\mathbf{T}_1^-,\alpha}]_{l=1,f=1;l=1,f=1} &= [P_{N}]_{l=1,m;l=1,m}  \nonumber\\
                                                          &= C_3(N)~~~~\forall \alpha,m \,,
	\end{align}
	we have the same combination coefficients
	\begin{equation}
		[G_{l;\Gamma}]_{N',F'=1;N,F=1} = \sqrt{C_3(N')\, C_3(N)} \,,
	\end{equation}
	for the pure $s$-wave and pure $p$-wave cases.
	So we just reproduce the result in \refref{Wu:2014vma}. 

        In fact, there can be no state in some $N$, e.g., we know $C_3(7)=0$. If these states are
        not discarded, there will be spurious states in the spectrum, which are not physical.

	\subsubsection*{Splitting of the $d$-wave}

	Here only $N=1$ ($6$ states) and $N=9$ ($30$ states) will be considered, then we have
	\begin{align}
		[G_{l=2;\mathbf{E}^+,\alpha=1}]_{f=1;N=1,F=1} &= \sqrt{15} \,,\nn\\
		[G_{l=2;\mathbf{E}^+,\alpha=1}]_{f=1;N=9,F=1} &= \sqrt{\frac{65}{3}} \,,\nn\\
		\text{Discarded: }[G_{l=2;\mathbf{T}_2^+,\alpha=1}]_{f=1;N=1,F=1} &= 0 \,,\nn\\
		[G_{l=2;\mathbf{T}_2^+,\alpha=1}]_{f=1;N=9,F=1} &= \frac{\sqrt{320}}{3} \,.
	\end{align}
	So the matrix form of the reduced $\hat V_L$ in $\mathbf{E}^+$ will be
	\begin{equation}
		\frac{1}{4\pi L^3}\,
		\begin{pmatrix}
		15\, v_2(k_1,k_1) & 5 \sqrt{13}\, v_2(k_1,k_9) \\
		5 \sqrt{13}\, v_2(k_9,k_1) & \frac{65}{3}\, v_2(k_9,k_9) \\
		\end{pmatrix}
		\,,
	\end{equation}
	and the matrix form of the reduced $\hat V_L$ in $\mathbf{T}_2^+$ will be
	\begin{equation}
		\frac{1}{4\pi L^3}\,
		\begin{pmatrix}
		\frac{320}{9} v_2(k_9,k_9) \\
		\end{pmatrix}
		\,,
	\end{equation}
	where $v_l(k,k')$ is the $l$ partial wave potential and $k_i=2\pi \sqrt{i}/L$ is the discrete momentum.


\subsection{Case of $\mathbf{F_\text{cut}(\Gamma,{l_\text{cut}})=1+1}$}
For the case \( F_{cut}(\Gamma,l_{cut}) = 1+1\,, \) \( f \) is always one, and therefore there are two states \( \ket{N;l_1,\Gamma,f=1,\alpha} \) and \( \ket{N;l_2,\Gamma,f=1,\alpha}, \) for some fixed \( N, \alpha\,. \)
We follow the same notation introduced around \eref{eq:rPN}, which is suppressing the indices $N,~\Gamma,~\alpha$ and $f$, and using $\ket{l_i}$ and $\ket{F_i}$ to represent $\ket{N;l_i,\Gamma,f=1,\alpha}$ and $\ket{N;\Gamma,F_i,\alpha}$ respectively. For completeness, we reintroduce the reduced P-Matrix \eref{eq:rPN}
\begin{equation}
  \tilde{P}_{N;\Gamma,\alpha} =
  \begin{pmatrix}
    \braket{l_1|l_1} & \braket{l_1|l_2} \\
    \braket{l_2|l_1} & \braket{l_2|l_2}
  \end{pmatrix}\,. \label{eq:Ptilde}
\end{equation}
To calculate the combination coefficients $\left[M_{l_i;\Gamma,\alpha}\right]_{f=1;N,F_j}=\braket{l_i|F_j}$, the states \( \ket{l_i} \) must be transformed to the new, orthonormal basis in \( \ket{F_i}\,. \)
This orthonormalisation can be performed in a number of ways, one is the eigenmode-based method introduced around \eref{eq:rPN}, another is the Gram-Schmidt process.

For this process, the projection of \( \ket{Y} \) onto the span of \( \ket{X} \) is defined as
\begin{equation}
  \text{proj}_{\ket{X}}\left(\ket{Y}\right) = \left\{
      \begin{array}{lr}
        0\,, & \ket{X} = 0 \\
        \frac{\braket{Y|X}}{\braket{X|X}}\, \ket{X}\,, & \ket{X} \neq 0
      \end{array} \right. \,.
\end{equation}
Defining some \( \ket{\tilde{F}}\,, \) where \( \ket{F} = \ket{\tilde{F}} / \sqrt{\braket{\tilde{F}|\tilde{F}}}\,, \) the Gram-Schmidt process gives
\begin{align}
  \ket{\tilde{F}_1} &= \ket{l_1}\,, \\
  \ket{\tilde{F}_2} &= \ket{l_2} - \text{proj}_{\ket{\tilde{F}_1}}\left(\ket{l_2}\right)\,.
\end{align}
Normalising these states requires careful consideration of the cases where \( \ket{l_1} = 0 \) and \( \ket{l_2} = 0\,. \)
When the original states \( \ket{l_i} \) are non-zero, \( \ket{\tilde{F}_i} \) must be normalised to \( \ket{F_i} = \ket{\tilde{F}_i} / \braket{\tilde{F}_i|\tilde{F}_i}\,. \)
These normalisation factors \( \braket{\tilde{F}_1|\tilde{F}_i} \) and \( \braket{\tilde{F}_2|\tilde{F}_2} \) are given by
\begin{align}
  \braket{\tilde{F}_1|\tilde{F}_1} &= \braket{l_1|l_1}\,, \\
  \braket{\tilde{F}_2|\tilde{F}_2} &= \braket{l_2|l_2} - \frac{\braket{l_2|l_1}^2}{\braket{l_1|l_1}}\,, \nonumber\\
                       &= \frac{1}{\braket{l_1|l_1}}\left( \braket{l_1|l_1}\,\braket{l_2|l_2} - \braket{l_1|l_2}^2  \right)\,, \nonumber\\
                       &= \frac{\det(\tilde{P}_{N;\Gamma\alpha})}{\braket{l_1|l_1}}\,.
\end{align}
Therefore the final, orthonormal states are given as
\begin{align}
  \ket{F_1} &= \left\{
      \begin{array}{lr}
        0\,, & \ket{l_1} = 0 \\
        \frac{\ket{l_1}}{\sqrt{\braket{l_1|l_1}}}\,, & \ket{l_1} \neq 0
      \end{array} \right. \,, \label{eq:F1}\\
  \ket{F_2} &= \left\{
      \begin{array}{lr}
        0\,, & \ket{l_1} = \ket{l_2} = 0 \\
        \frac{\ket{l_2}}{\sqrt{\braket{l_2|l_2}}}\,, & \ket{l_1} = 0\,, \ket{l_2} \neq 0 \\
        \frac{ \ket{l_2} - \frac{\braket{l_2|l_1}}{\braket{l_1|l_1}} \,\ket{l_1} }{ \sqrt{\det{\tilde{P}_{N;\Gamma,\alpha}} / \braket{l_1|l_1}} }\,, & \ket{l_1} \neq 0\,, \det(\tilde{P}_{N;\Gamma,\alpha}) \neq 0
      \end{array} \right. \,. \label{eq:F2}
\end{align}
From these expressions of \( \ket{F_i}\,, \) we see that there are three possible values for \( F_{max}\,. \) When \( \ket{F_1} \) and \( \ket{F_2} \) both vanish, we have that \( F_{max} = 0\,. \) \( F_{max} = 1 \) when only one \( \ket{F_i} \) vanishes but the other does not, and \( F_{max} = 2 \) for only the second entry of \eref{eq:F1} and the third entry of \eref{eq:F2}.
As these states are now constructed in terms of only \( \ket{l_i} \) as given in the elements of \( \tilde{P}_{N;\Gamma\alpha} \) from \eref{eq:SFV-defPG}, it is therefore simple to calculate the matrix defined in \eref{eq:SFV-defGS},
\begin{align}
  \left[M_{l;\Gamma,\alpha}\right]_{f;N,F} &= \braket{N,l_j;\Gamma,f,\alpha|N;\Gamma,F_i,\alpha}\,, \nonumber\\
                                                   &= \braket{l_j|F_i}\,.
\end{align}

	\subsubsection*{Mixing between s-wave and g-wave}

	Here only $N=1$ ($6$ states) and $N=9$ ($30$ states) will be considered as before, we have
	\begin{equation}
		\tilde{P}_{N=1;\mathbf{A}_1^+,\alpha=1} =
		\begin{pmatrix}
		6 & 3\sqrt{21} \\
		3\sqrt{21} & \frac{63}{2} \\
		\end{pmatrix}
		\,,
	\end{equation}
	and
	\begin{equation}
		\tilde{P}_{N=9;\mathbf{A}_1^+,\alpha=1} =
		\begin{pmatrix}
		30 & -\frac{25}{3}\sqrt{\frac{7}{3}} \\
		-\frac{25}{3}\sqrt{\frac{7}{3}} & \frac{9835}{162} \\
		\end{pmatrix}
		\,.
	\end{equation}
	The determinants give
	\begin{equation}
		\det(\tilde{P}_{N=1;\mathbf{A}_1^+,\alpha=1}) = 0~~,~~\det(\tilde{P}_{N=9;\mathbf{A}_1^+,\alpha=1}) = \frac{44800}{27} \,,
	\end{equation}
	so there is only one state in $N=1$, while there are two states in $N=9$. Then we have (we use $[G_{l}]_{N,F}$ to represent $[G_{l;\Gamma=\mathbf{A}_1^+,\alpha=1}]_{f=1;N,F}$)
	\begin{align}
		&[G_{l=0}]_{N=1,F=1} = \sqrt{6}\,,~[G_{l=0}]_{N=9,F=1} = \sqrt{30}\,, \nn\\
		&[G_{l=0}]_{N=9,F=2} = 0\,,~[G_{l=4}]_{N=1,F=1} = 3\sqrt{\frac{7}{2}}\,,\nn\\
		&[G_{l=4}]_{N=9,F=1} = -\frac{5}{9}\sqrt{\frac{35}{2}}\,,~[G_{l=4}]_{N=9,F=2} = \frac{8}{9}\sqrt{70} \,.
	\end{align}
        %
        %
	So the matrix form of the reduced $\hat V_L$ in $\mathbf{A}_1^+$ will be
	\begin{align}
	&\frac{1}{4\pi L^3} \times \nn\\
	&\begin{pmatrix}
	 \frac{12 \,v_0^{1,1}+63 \,v_4^{1,1}}{2} & \frac{\sqrt{5} (36 \,v_0^{1,9}-35 \,v_4^{1,9})}{6} & \frac{56\sqrt{5} \,v_4^{1,9}}{3}  \\
	 \frac{\sqrt{5} (36 \,v_0^{9,1}-35 \,v_4^{9,1})}{6} & \frac{5(972\,v_0^{9,9}+175\,v_4^{9,9})}{162} & \frac{-1400\,v_4^{9,9}}{81}  \\
	 \frac{56\sqrt{5} \,v_4^{9,1}}{3}  & \frac{-1400\,v_4^{9,9}}{81}  & \frac{4480\,v_4^{9,9}}{81}  \\
	\end{pmatrix}
	\,,
	\end{align}
	where $v_l^{i,j}=v_l(k_i,k_j)\,.$


\section{Characteristics of the $\mathbf{P}$-Matrix}\label{sec:CPM}

\begin{figure*}[h]
\centering
\includegraphics[width=\textwidth]{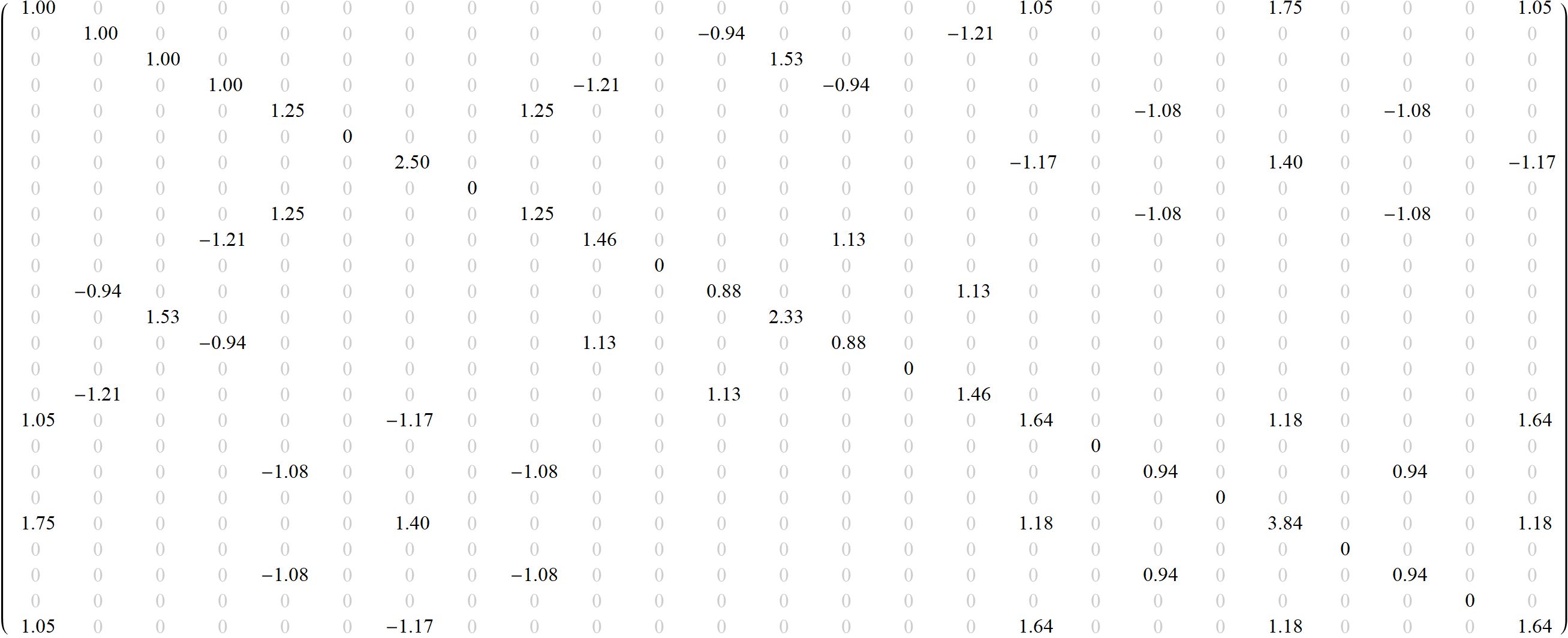},
\caption{$P_{1}/C_3(1)$ with $C_3(1)=6$: $25\times25$ matrix ordered as $(l,m) =
  (0,0),\,(1,-1),\,(1,0),\,(1,1),\,\cdots,\,(4,4).$ As only one {\it shell} is contributing,
  off-diagonal elements can be the same order of magnitude as the diagonal elements.}
\label{fig:P1}
\end{figure*}

\begin{figure*}[h]
\centering
\includegraphics[width=\textwidth]{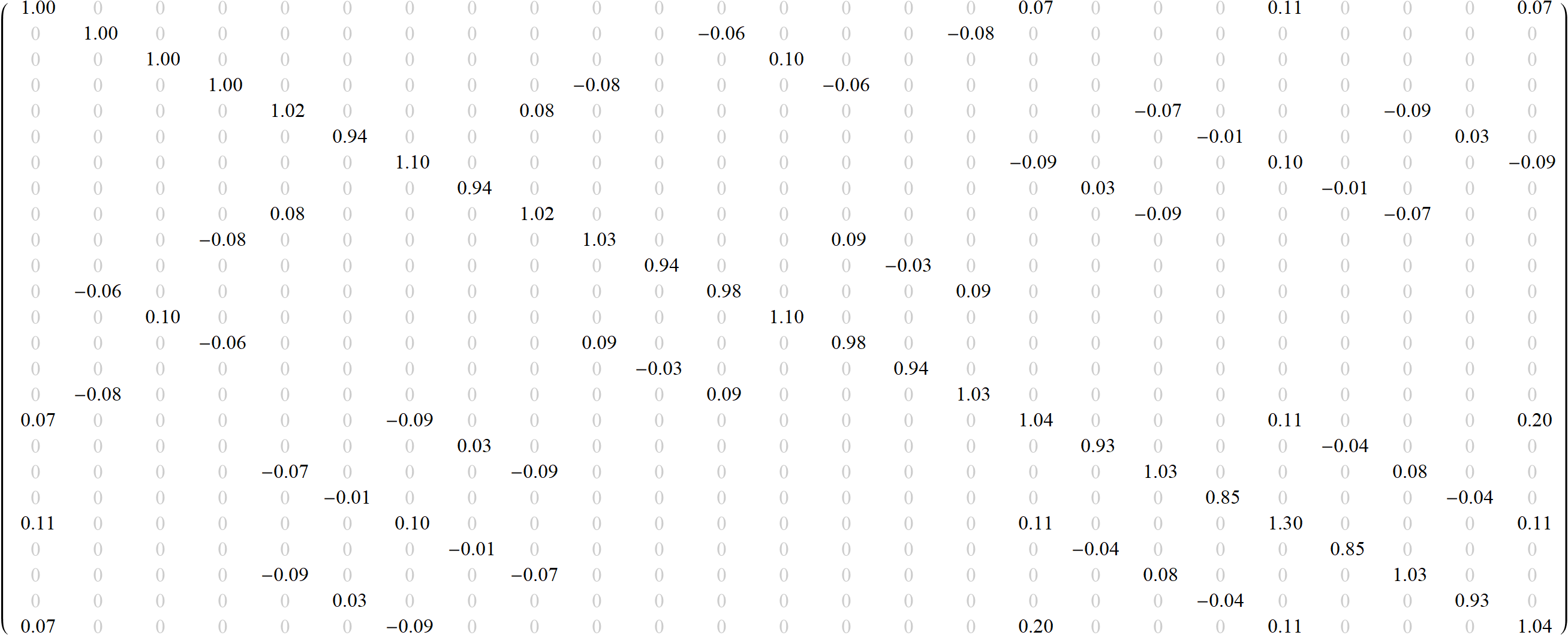}
\caption{$P_{581}/C_3(581)$ with $C_3(581)=336$: $25\times25$ matrix ordered as $(l,m) =
  (0,0),\,(1,-1),\,(1,0),\,(1,1),\,\cdots,\,(4,4)$.  Here the contributions of several {\it shells}
  provide an opportunity for cancellations in the evaluation of the $P$ matrix and the matrix
  approaches the identity.} 
\label{fig:P581}
\end{figure*}

\begin{figure*}[h]
\centering
\includegraphics[width=\textwidth]{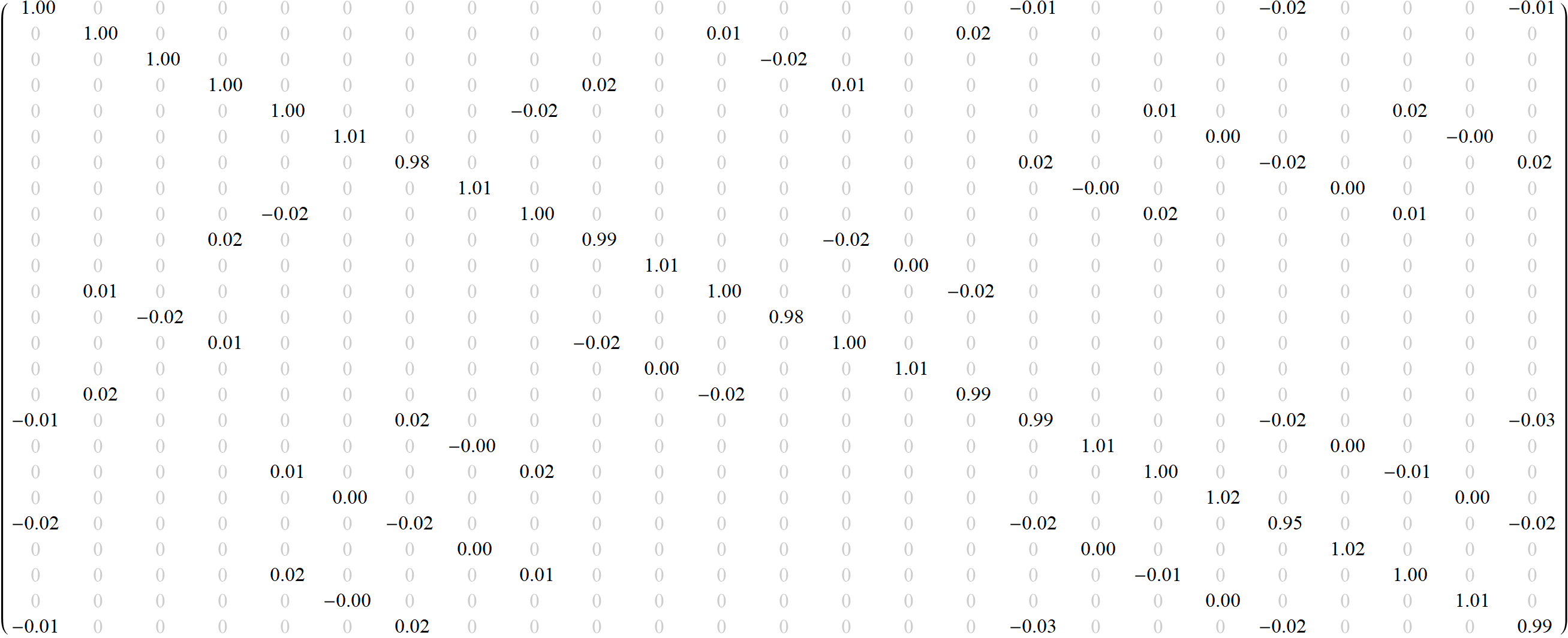}
\caption{$P_{941}/C_3(941)$ with $C_3(941)=552$: $25\times25$ matrix ordered as $(l,m) =
  (0,0),\,(1,-1),\,(1,0),\,(1,1),\,\cdots,\,(4,4).$ In this case $C_3(N)$ takes the very large value
  of 552, reflecting the contributions of many {\it shells}.  Averaging over these shells enables
  this $P$ matrix to approximate the identity.  However the energy associated with $N=941$ is very
  large on contemporary lattices.}
\label{fig:P941}
\end{figure*}

The $P$-matrix is calculated via the definition of \eref{eq:SFV-defP}.  Here we provide a few
examples of the numerical values for $P_N = P_1$, $P_{581}$ and $P_{941}$ in Figs.~\ref{fig:P1},
\ref{fig:P581}, and \ref{fig:P941} respectively.  The off-diagonal nature of the matrices is a
direct illustration of partial-wave mixing in the periodic finite-volume of the cubic group.

The values of $N$ selected provide an overview of the characteristic properties of $P_N$.  $P_1$ is
characteristic of low values of $N$ where off-diagonal elements can be the same order of magnitude
as the diagonal elements.  The values of $N=581$ and $941$ illustrate moderate and large levels of
shell mixing which can aid in suppressing off-diagonal elements.  As $N$ grows, the opportunity for
shell mixing increases and the P-Matrix will approach the identity as discussed in \sref{sec:MPW}.
Also, we note the P-Matrix of the cubic-group basis defined in \eref{eq:SFV-defPG} also respects
this behavior, as the unitary transformation of an identity matrix is still an identity matrix.

\end{appendix}

	\newpage
	\bibliography{refs}

\end{document}